\documentclass[twocolumn]{aastex62}

\usepackage{subfigure}
\usepackage{url}
\usepackage{hyperref}
\usepackage{longtable}
\usepackage{natbib}
\usepackage{amsmath}
\usepackage{listings}
\usepackage[normalem]{ulem}
\usepackage{bm}
\usepackage{comment}
\usepackage{array}
\newcolumntype{P}[1]{>{\centering\arraybackslash}p{#1}}
\newcolumntype{M}[1]{>{\centering\arraybackslash}m{#1}}

\usepackage{xcolor, fontawesome}
\definecolor{twitterblue}{RGB}{64,153,255}
\newcommand{\twitter}[1]{\href{https://twitter.com/#1 }{\textcolor{twitterblue}{\faTwitter}\,\tt \textcolor{twitterblue}{@#1}}}

\definecolor{Code}{rgb}{0,0,0}
\definecolor{Decorators}{rgb}{0.5,0.5,0.5}
\definecolor{Numbers}{rgb}{0.5,0,0}
\definecolor{MatchingBrackets}{rgb}{0.25,0.5,0.5}
\definecolor{Keywords}{rgb}{1,0,0}
\definecolor{self}{rgb}{0,0,0}
\definecolor{Strings}{rgb}{0,0.63,0}
\definecolor{Comments}{rgb}{0,0.63,1}
\definecolor{Backquotes}{rgb}{0,0,0}
\definecolor{Classname}{rgb}{0,0,0}
\definecolor{FunctionName}{rgb}{0,0,0}
\definecolor{Operators}{rgb}{0,0,0}
\definecolor{Background}{rgb}{0.98,0.98,0.98}
\definecolor{Booleans}{rgb}{0.572,0,0.572}
\definecolor{BuiltinFunction}{rgb}{0.572,0,0.572}
\definecolor{BuiltinConstant}{rgb}{0.572,0,0.572}
\definecolor{Asterisk}{rgb}{0.670,0,1}

\lstdefinelanguage{Python}{
    	numbers=left,
    	numberstyle=\footnotesize,
    	numbersep=7pt,
    	xleftmargin=1.26em,
    	framextopmargin=2em,
    	framexbottommargin=2em,
    	showspaces=false,
    	showtabs=false,
    	showstringspaces=false,
    	frame=l,
    	tabsize=4,
    	stepnumber=1,
	basicstyle=\small\ttfamily,
    	backgroundcolor=\color{Background},
	stringstyle=\ttfamily\color{Strings},
	morekeywords={import,from,class,def,while,if,in,elif,else,not,or,print,break,continue,return,access,as,except,exec,finally,global,import,lambda,pass,print,raise,try,assert},
    	keywordstyle={\color{Keywords}\bfseries}, 
	otherkeywords={[2]*},
	keywordstyle={[2]\color{Asterisk}},
}

\usepackage{color}

\newcommand{\shrug}{\texttt{\raisebox{0.75em}{\char`\_}\char`\\\char`\_\kern-0.5ex(\kern-0.25ex\raisebox{0.25ex}{\rotatebox{45}{\raisebox{-.75ex}"\kern-1.5ex\rotatebox{-90})}}\kern-0.5ex)\kern-0.5ex\char`\_/\raisebox{0.75em}{\char`\_}}}

\newcommand{\kep}{{\it Kepler}}
\newcommand{\ktwo}{{\it K2}}
\newcommand{\tess}{{\it TESS}}

\newcommand{\fulltess}{{\it Transiting Exoplanet Survey Satellite}}

\newcommand{\eleanor}{\texttt{eleanor}}

\newcommand{\chicago}{Department of Astronomy and Astrophysics, University of
Chicago, 5640 S. Ellis Ave, Chicago, IL 60637, USA}
\newcommand{\sagan}{Sagan Fellow}

\newcommand{\ufcg}{Universidade Federal de Campina Grande, R. Aprígio Veloso, 882 Universitário, Campina Grande, 58429, Brazil}
\newcommand{\ames}{NASA Ames Research Center, Moffett Field, CA 94035, USA}

\newcommand{\flatiron}{Center for Computational Astrophysics, Flatiron Institute, 162 Fifth Ave, New York, NY 10010, USA}

\newcommand{\baeri}{Bay Area Environmental Research Institute, 625 2nd St Ste. 209, Petaluma, CA 94952}

\newcommand{\ipac}{Caltech/IPAC-NASA Exoplanet Science Institute, M/S 100-22, 770 S. Wilson Ave, Pasadena, CA 91106 USA}
\newcommand{\duke}{Department of Physics, Duke University, 120 Science Drive, Durham, NC, 27708, USA}

\begin{document}
\title{\eleanor: An open-source tool for extracting light curves from the \tess\ Full-Frame Images}

\shorttitle{\eleanor\ Light curves} 
\shortauthors{Feinstein et al.}

\author[0000-0002-9464-8101]{Adina~D.~Feinstein}
\affiliation{\chicago}

\author[0000-0001-7516-8308]{Benjamin~T.~Montet}
\altaffiliation{\sagan}
\affiliation{\chicago}

\author[0000-0002-9328-5652]{Daniel Foreman-Mackey}
\affiliation{\flatiron}

\author[0000-0001-9907-7742]{Megan E. Bedell}
\affiliation{\flatiron}

\author[0000-0003-2657-3889]{Nicholas Saunders}
\affiliation{\ames}
\affiliation{\baeri}

\author[0000-0003-4733-6532]{Jacob L. Bean}
\affiliation{\chicago}

\author[0000-0002-8035-4778]{Jessie L. Christiansen}
\affiliation{\ipac}

\author[0000-0002-3385-8391]{Christina Hedges}
\affiliation{\ames}
\affiliation{\baeri}

\author[0000-0002-0296-3826]{Rodrigo Luger}
\affiliation{\flatiron}

\author[0000-0002-4934-5849	]{Daniel Scolnic}
\affiliation{\duke}

\author{Jos\'e Vin\'icius de Miranda Cardoso}
\affiliation{\ufcg}

\correspondingauthor{Adina~D.~Feinstein; \twitter{afeinstein20}}
\email{afeinstein@uchicago.edu}


\begin{abstract}

During its two year prime mission the \fulltess\ (\tess) will perform a time-series photometric survey covering over 80\% of the sky. This survey comprises observations of 26 24$^\circ \times 96^\circ$ sectors that are each monitored continuously for approximately 27 days. The main goal of \tess\ is to find transiting planets around 200,000 pre-selected stars for which fixed aperture photometry is recorded every two minutes. However, \tess\ is also recording and delivering Full-Frame Images (FFIs) of each detector at a 30 minute cadence. We have created an open-source tool, \eleanor, to produce light curves for objects in the \tess\ FFIs. Here, we describe the methods used in \eleanor\ to produce light curves that are optimized for planet searches. The tool performs background subtraction, aperture and PSF photometry, decorrelation of instrument systematics, and cotrending using principal component analysis. We recover known transiting exoplanets in the FFIs to validate the pipeline and perform a limited search for new planet candidates in Sector 1. Our tests indicate that \eleanor\ produces light curves with significantly less scatter than other tools that have been used in the literature. Cadence-stacked images, and raw and detrended \eleanor\ light curves for each analyzed star will be hosted on MAST, with planet candidates on ExoFOP-TESS as Community \tess\ Objects of Interest (CTOIs). This work confirms the promise that the \tess\ FFIs will enable the detection of thousands of new exoplanets and a broad range of time domain astrophysics. 
\end{abstract}


\section{Introduction} \label{sec:intro}

The recently retired \kep\ and \ktwo\ missions \citep[]{Borucki10, Howell14} revealed tremendous new insight into the frequency and architectures of exoplanetary systems. It is therefore timely that \tess, the \fulltess\ \citep[]{Ricker15}, is already detecting new transiting planets \citep{clouteir18, dragomir19, esposito18, gandolfi18,gunther19, Huang18, huber19, jones18, Kostov19, quinn19, rodriguez19, trifonov19, vanderspek19, wang19}.

The \tess\ prime mission is a two year survey observing roughly 80\% of the sky for exoplanet transits.  \tess's four cameras are aligned along a 96$^\circ \times 24^\circ$ degree sector of the sky and are observed for approximately 27 days. There are 20,000 stars pre-selected by TESS mission operators and through Guest Investigator (GI) proposals observed every sector in a short 2-minute cadence mode. These targets have been selected not only for new exoplanet candidate searches, but also for asteroseismic studies, Solar System research, and additional galactic and extragalactic astrophysics.\footnote{See https://heasarc.gsfc.nasa.gov/docs/tess/approved-programs.html for examples}

In addition to the short cadence photometry of the main targets, \tess\ obtains images, known as Full-Frame Images (FFIs), of each sector at 30-minute cadence. There are roughly one million stars in the FFIs brighter than I=16 mag in each sector of observations. As such, the FFIs provide a huge data mining archive for many different areas of astronomy, including the search for new transiting exoplanet candidates. Simulations from \cite{Barclay18} predict that within the FFIs there will be an additional 3,100 detectable exoplanets orbiting bright (T\textsubscript{mag} $<$ 11.0) stars, and a further 10,000 detectable exoplanets orbiting fainter stars. Within the FFIs, the authors project $\sim$ 1,500 large planets (R$_{p}$\,$>$\,4\,R$_\oplus$) and $\sim$ 400 small planets (R$_{p}$\,$\leq$\,4\,R$_\oplus$) will be identified, with 67\% of the planets orbiting F and G type stars \citep{Barclay18}.

Asteroseismology, the study of stellar oscillations to probe the internal structure of stars, will additionally benefit from the \tess\ FFIs. The ability to measure stellar oscillations with long cadence data has been previously demonstrated with the \kep\ 30-minute cadence data \citep{hekker17}. There is also the possibility to study Solar System objects, and galactic and extragalactic sources using the FFI data. Using a difference imaging approach and \ktwo\ long-cadence data, \cite{dimitri18, shappee18} were able to obtain a light curve for a supernova, SN2018oh, roughly 52.7\,Mpc away. The light curve begins 18 days before peak brightness, which is a feat that cannot be achieved by even the most advanced surveys that are triggered by supernova events. In addition to supernovae, teams such as \cite{molnar15} were able to detect extragalactic RR Lyrae stars in Leo IV, a dwarf galaxy at a distance of $\sim$ 154\,kpc. Using \ktwo\ observations, they observed the farthest measurement of the Blazhko effect, or long-period modulations in the period and amplitude of the light curve. This was the first discovery of the effect outside of the Milky Way and the Magellanic Clouds.

\begin{figure}[!ht]
\begin{center}
\includegraphics[width=0.45\textwidth,trim={0.25cm 0 0 0}]{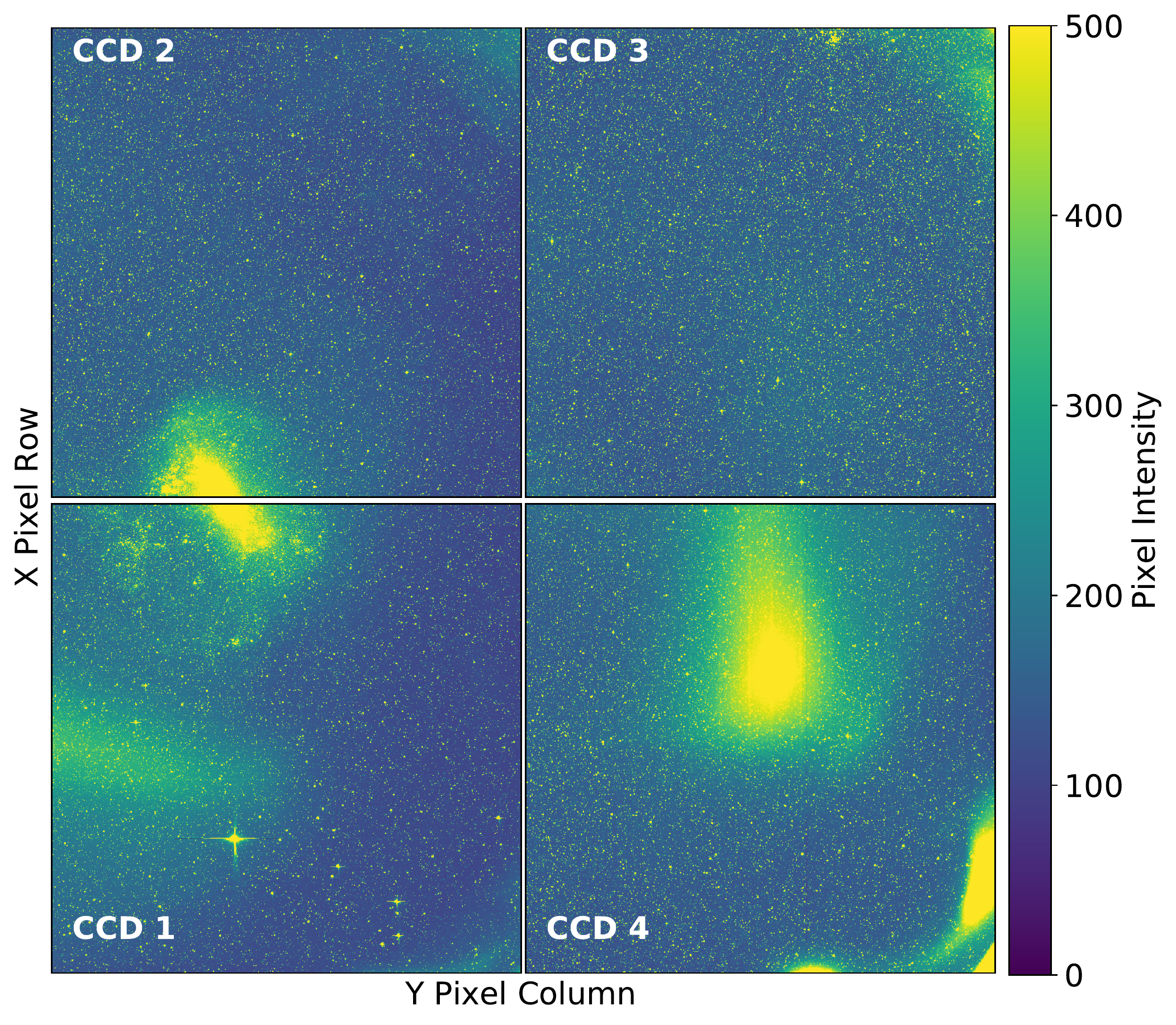}
\caption{An example FFI from Sector 1, Camera 4. There is noticeable structured background in the corner of the CCDs 3 and 4 as well as in the center of CCDs 2 and 4. The Large Magellanic Cloud is seen in CCDs 1 and 2. \label{fig:ffi}}
\end{center}
\end{figure}

Despite their substantial scientific potential, there is significant processing that needs to be completed before extracting usable light curves from the FFIs. A background correction can be approximated over the entire FFI, however, it would not properly account for regions with more localized issues (see Figure~\ref{fig:ffi}). On the FFI scale, systematic effects can overwhelm astrophysical signals, especially when the telescope is near perigee. Additionally, the FFIs are not in a user-friendly format. Each FFI is $\sim$35 MB. In order to complete photometry for a single target in a single sector, the user must have access to $\sim$45 GB of storage for any given sector and 1 TB for the entire Southern Hemisphere. This makes it challenging for users without vast computational resources to fully exploit the FFIs. 

The recently ended \ktwo\ mission motivated the creation of several community driven pipelines for data reduction. There is significant benefit to having multiple pipelines with different methods for data reduction for the same data sets. For example, one could find a new planet candidate in the \texttt{EVEREST} light curves \citep[]{Luger18} and compare to the \texttt{K2SFF} light curves \citep[]{Vanderburg14} to determine if the signal is real. Additionally, \cite{Shaya15} created the \kep\ Extra-Galactic Survey (KEGS), with the goal of producing light curves for extra-galactic sources. These pipelines were especially useful due to the large systematics found within the \ktwo\ data. Other pipelines, including \texttt{K2VARCAT} \citep[]{Armstrong15a}, \texttt{K2SC} \citep[]{Aigrain15}, \texttt{POLAR} \citep[]{Barros15}, and \texttt{K2P\textsuperscript{2}} \citep[]{Lund15} created light curves for the public to use as well, each with their own methods for removing systematics, and therefore their own strengths and weaknesses.

In this article, we present the \eleanor\ pipeline\footnote{https://github.com/afeinstein20/eleanor} for light curve extraction from the \tess\ FFIs and publicly available \eleanor\ light curve data products. In Section \ref{sec:methods}, we describe the methods used to create our light curves. In Section \ref{sec:results}, we demonstrate the photometric capabilities of \eleanor\ by presenting early science results, including recovering known transiting planets, new planet candidates, and other stellar variability. In Section \ref{sec:availability}, we discuss the light curve data products and their availability, and how to download the open-source software package.

\section{Creating Light Curves}\label{sec:methods}

In this section, we describe how \eleanor\ extracts light curves from the FFIs. We first create a pointing model and assign quality flags on the FFI level. Then, we cut out intermediate ``postcards" (148 $\times$ 104 pixels) that are time-stacked and background-subtracted. The Target Pixel Files (TPFs; 13 $\times$ 13 pixels) are extracted from the postcards. \eleanor\ tests multiple apertures to find the best light curve for transiting exoplanet searches for each target. The TPF pixel time-stacked cut-out and light curves are stored in the \eleanor\ data products, which are described in \ref{subsec:products}.

\subsection{Pointing Model}\label{subsec:PM}

We first download all of the FFIs for a given sector. Within the FFIs, we build a pointing model to ensure a true position of the star on the detector. Due to spacecraft motion, the World Coordinate System (WCS) written in the headers of the FFIs may not return an accurate transformation from pixel space to sky position. For each CCD, we complete a search of targets in the \tess\ Input Catalog \cite[TIC, version 7.0;][]{stassun18} with 7.5 $\leq$ $T$\textsubscript{mag} $\leq$ 12.5 and 0 $\leq$ contamination $\leq$ 5 $\times$ 10\textsuperscript{-3}, leaving us with bright, but not saturated, uncrowded stars to calibrate our pointing model. 

We then determine the affine transformation that minimizes the square of the differences between the predicted and observed positions for all the stars that met our search criteria on the detector at each cadence. The affine transformation accounts for any difference between the predicted and observed stellar positions, such as a rotation or translation of the position of the spacecraft, or changes in apparent position of the stars due to differential velocity aberration. The predicted position of the stars are determined using the RA and Dec from the TIC and the WCS solution given in the FFIs. Corrected pixel positions, using the pointing model, of the stars are saved in the \eleanor\ data product.

\subsection{From FFIs to Postcards}\label{subsec:postcards}

Before extracting light curves, we create intermediate data products called ``postcards" that represent a more efficient format for analyzing single targets with FFI data. Postcards are 148 $\times$ 104 pixel cut-out regions of the FFIs, and are created with a 50 pixel overlap between each postcard to avoid edge effects for individual stars. Unlike the FFIs, the postcards are time-stacked, including all cadences for which observations are available, and are background-subtracted.

As the background of the FFIs is highly structured and varies greatly across the detector (Figure \ref{fig:ffi}), the more localized scaling of the postcards provides a sufficient region for initial backgroud subtraction. We use a constant background on the postcard level with the \texttt{photutils} function \texttt{MMMBackground}, which calculates a background of
\begin{equation}
    background = 3*mean - 2*median
\end{equation}
for each cadence in the postcard. We tested several combinations of the mean and median pixel values, as given in the \texttt{photuils} background functions across each cadence and concluded this relation minimized the scatter of the background pixels. TPFs and light curves are extracted from the background-subtracted postcards.

We also model a two-dimensional background across each postcard. On each postcard, we stack frames across all cadences with no quality flags set (as described in Section \ref{subsec:flags}) and identify the bottom 40\% of pixels in flux, removing the extended PSF of bright stars across the postcard. We then take these $\approx$ 6300 pixels and apply PCA to identify common systematics shared between each of these. We build a set of vectors, considering the 5 most significant modes, and shifting them by up to 15 cadences in order to capture variability that occurs at different times across the postcard. We then build a linear model of these vectors that minimizes the scatter in observed brightness at each pixel. We remove pixels that are on average more than 1-sigma discrepant from the mean value at each cadence, which effectively removes pixels in streaks that are affected by bright asteroids. 
We then linearly interpolate these vectors across the entire postcard to model the expected background behind the bright stars, applying a low-pass filter to remove any signals in the background data with frequencies faster than four hours.

Within each postcard, the WCS from each FFI is conserved. The postcards also contain quality flags to highlight potentially corrupted cadences. We follow a two-step process for assigning the quality flags.

\subsection{Quality Flags}\label{subsec:flags}

The \tess\ mission assigns twelve different quality flags, eight of which are applicable to the FFIs. The quality issues are: attitude tweaks; the spacecraft is in coarse point; the spacecraft is in Earth point; an argabrightening event occurs; reaction wheel desaturation event occurs; a cosmic ray is detected on a collateral pixel row or column; there is stray light from the Earth or Moon in the camera field-of-view; or a ``manual exlude" set in the processing of short cadence data \citep[for more information see Table 28 in][]{tenenbaum2018}. We copy these quality flags into our postcards by identifying short-cadence targets that fall on each camera-CCD pairing for a given sector. There are roughly 15 short-cadence observations for every FFI observation, and we follow the most conservative procedure: any applicable quality flag that falls during an individual FFI exposure is recorded in the postcards, using the same numeric identifiers for each individual quality flag as applied in the short-cadence data.

\begin{figure*}[!ht]
\begin{center}
\includegraphics[width=1.0\textwidth,trim={0.25cm 0 0 0}]{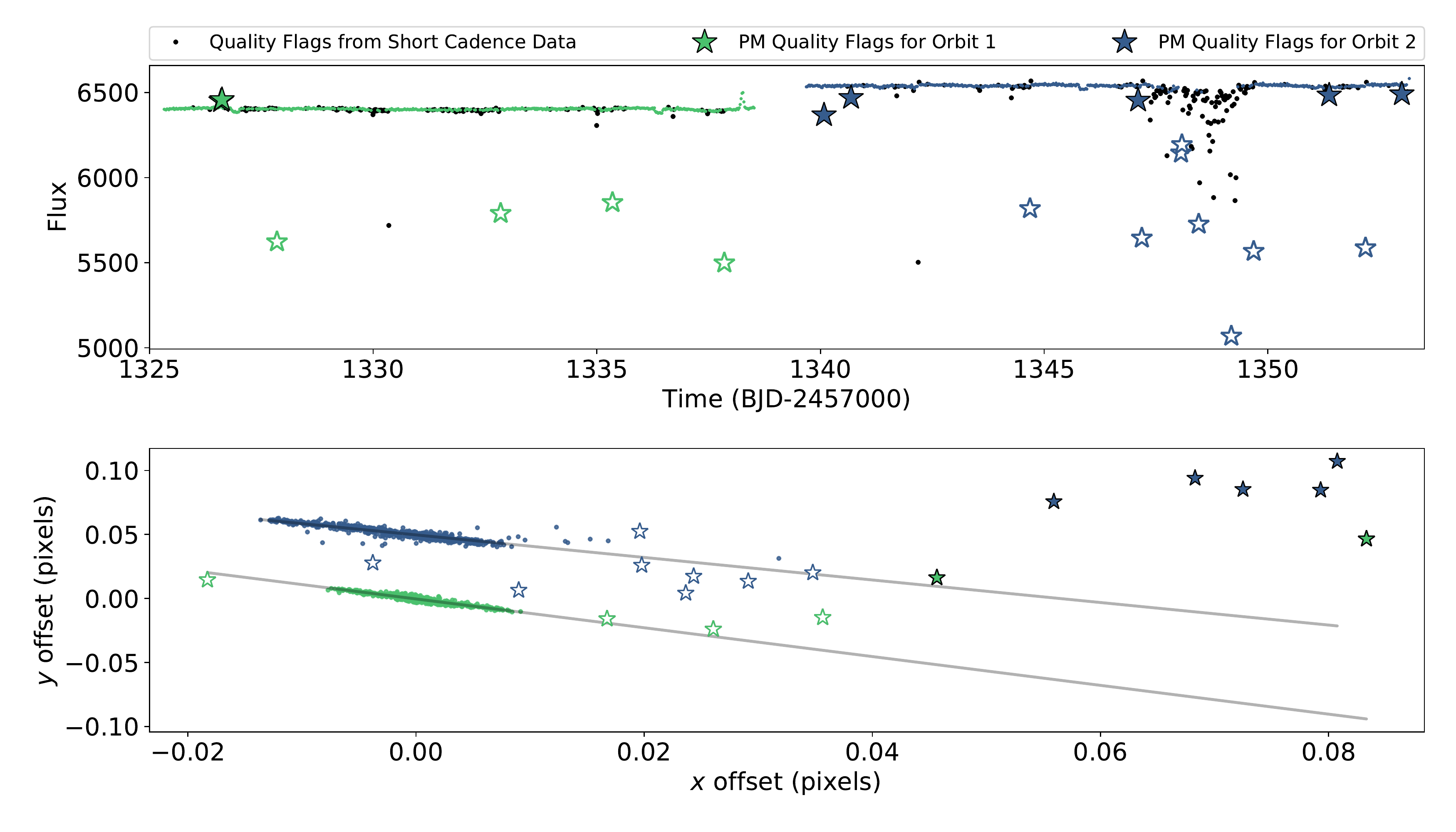}
\caption{An example of quality flags marked by significant shifts in $x$, $y$ pixel positions given by the pointing model (PM) for Sector 1. A line is fit to each orbit (corresponding to the different colors). Filled stars represent flagged data points offset from the PM from each orbit of the appropriate color. Open stars represent flagged data points from PM offset as well as flagged in the short cadence data. Black points represent quality flags taken from the short cadence data that apply to the entire FFI. There is some overlap between our marked quality flags and those from the short cadence data. \label{fig:pm_corr}}
\end{center}
\end{figure*}

Furthermore, we introduce our own quality flag based on the pointing model. We fit a line to the measured $x$ and $y$ pixel coordinates with the applied pointing model and complete an iterative sigma clipping at $2\sigma$, see Figure~\ref{fig:pm_corr}. The stars in Figure~\ref{fig:pm_corr} represent the bad pointing model cadences. We fit each orbit independently and assign a quality flag value of 4096. We apply the bad pointing model quality flag value to cadences that have quality flags from the short cadence data as well.

\subsection{From TPFs to Light curves} \label{subsec:lcs}

A Target Pixel File (TPF) is cut out for each target from the postcard. The default size for \eleanor\ TPFs is 13 $\times$ 13 pixels. The target is at the center of the TPF when possible. Photometry is completed on the TPF level. Due to the 50-pixel overlap between each postcard, targets may fall on multiple postcards; we extract the TPF and light curve from the postcard in which the target is closest to the center.

\subsubsection{Aperture Selection}

\begin{figure}[!ht]
\begin{center}
\includegraphics[width=0.4\textwidth,trim={0.25cm 0 0 0}]{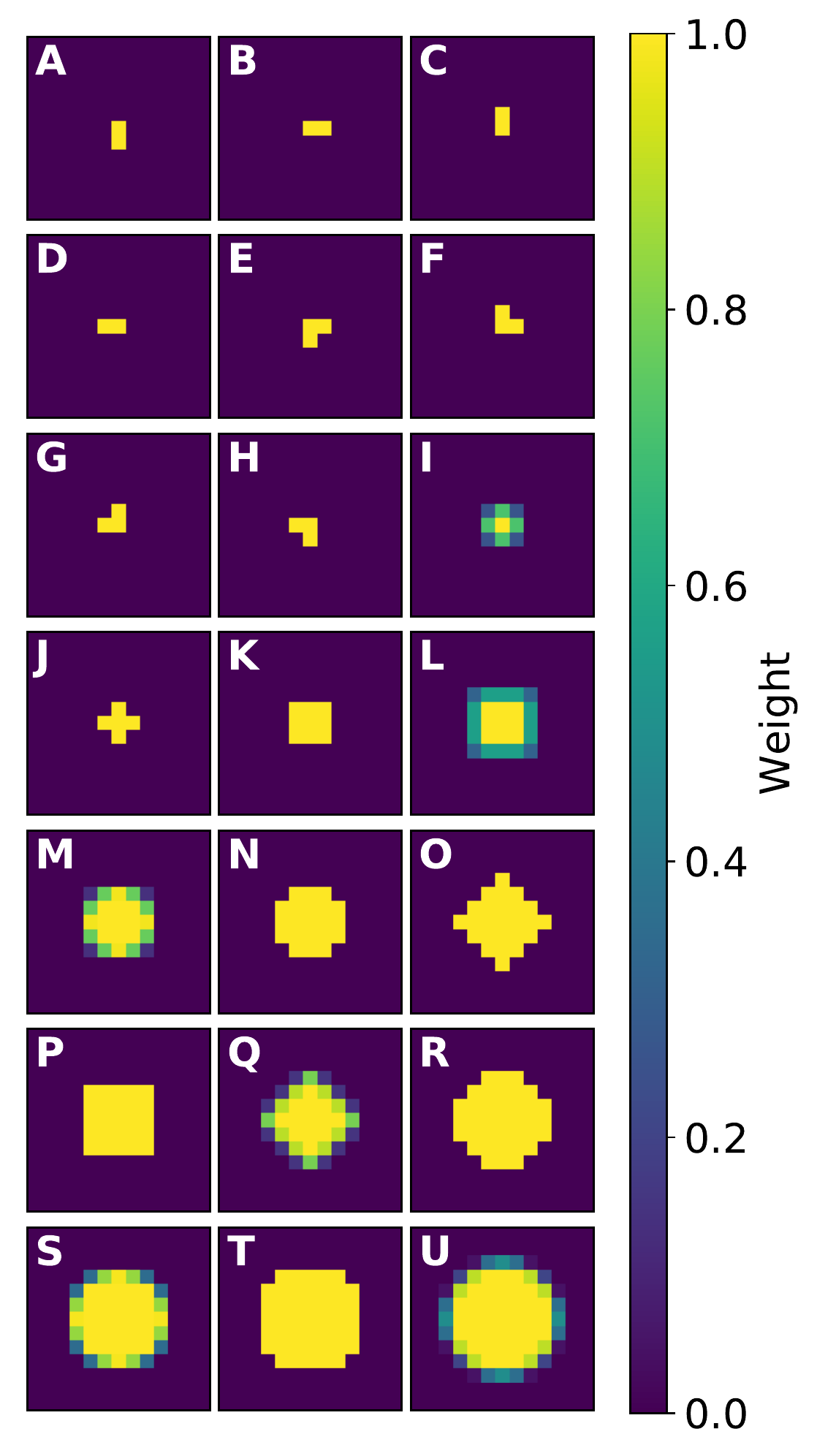}
\caption{The default library of apertures \eleanor\ uses to create light curves. We try a range of shapes and sizes as well as assigning varying weights to each pixel. The binary apertures have pixel values of 0 and 1; the weighted apertures have pixel values from 0 to 1. All apertures are applied and tested unless \texttt{crowded\_field = True}, in which case only apertures with aperture pixel sums less than 9 (apertures A-J) are used to extract light curves. \label{fig:aps}}
\end{center}
\end{figure}

Once the TPF has been extracted, \eleanor\ uses a pre-defined library of apertures of various shapes and sizes (see Figure \ref{fig:aps}) to measure photometry. We test apertures that were shown to work well for \kep\ photometric extraction, including 2 $\times$ 1 rectangles and an ‘L’ shape, both in four different orientations about the center. We also test standard circular and square apertures defined using the \texttt{photutils} package. We use both binary (pixel values of either 0 or 1) and weighted (pixel values ranging from 0 to 1) masks when extracting photometry. The weights for the non-binary apertures are determined by the exact fractional overlap of the aperture and each pixel, dictated by \texttt{photutils.aperture\_photometry()}. Circular apertures have radii of 1.25, 2.5, 3.5, and 4 pixels. Squares have lengths and widths of 3, 4.1, and 5 pixels, as well as a 3 pixel length and width that is rotated 45$^\circ$. We chose these orientations to maximize diversity in the aperture testing. All apertures and associated extracted light curves are saved in the \eleanor\ data product.

A user of the \eleanor\ software also has the option to define their own aperture masks as well, in the \texttt{eleanor.TargetData.get\_lightcurve()} function. Masks are required to be 2D arrays of the same length and height as the TPF. There is an additional option for users to customize a \texttt{photutils} aperture using \texttt{eleanor.TargetData.custom\_aperture()}. This is a function that allows the user to choose the radius or length and width and angle of a circular or rectangular aperture. Users can also define a position in TPF pixel space to offset the aperture from the center, where it is placed by default. The custom apertures can additionally be defined as binary or weighted.

The photometry is completed by multiplying the aperture and the TPF and summing the pixel product in each cadence. We define this as ``RAW\_FLUX". Note that the ``RAW\_FLUX" is background-subtracted as the pixels were extracted from a background-subtracted postcard. After the raw photometry for each aperture is extracted, we correct for possible systematic effects on an orbit-by-orbit basis, creating a flux time series called ``CORR\_FLUX". We regress the raw flux time series against a linear model of the $x$ pixel position, $y$ pixel position (both taken from our pointing model), measured background at the location of the TPF, and time, effectively removing any signals correlated with these parameters. We note that long timescale astrophysical signals, such as starspot-induced stellar variability for slowly rotating stars, or transients like supernovae could produce an approximately linear signal over a single orbit. This signal would then be removed by ``CORR\_FLUX'' in a similar way as it was removed by the \kep\ pipeline in the generation of their pre-search data conditioning (PDCSAP) flux. 

We test to see which background subtraction on the TPF level will increase the precision of the extracted light curve. The background routine is marked in the header by the flag ``BKG\_LVL", which reads either ``CONSTANT" indicating the background subtraction is using the \texttt{photutils.MMMBackground} routine, or ``2D\_BKG" indicating the two-dimensional modeled background was applied.

After a light curve has been extracted using every aperture shape and size and the light curve has received background treatment on both the postcard and postcard and TPF level, an ideal aperture is chosen for that target. Here, we are primarily interested in creating the best light curve possible for detecting planets, preserving the sharp features on short timescales induced by planet transits. In order to achieve this, we minimize the combined differential photometric precision (CDPP) of the light curve on one-hour timescales. The CDPP is a metric in units of parts-per-million (ppm) that was originally defined for \kep\ to assess the ability to detect a weak terrestrial planet transit signal in a light curve \citep[]{Christiansen12}. By definition, the CDPP is the root mean square of the photometric noise on transit timescales \citep[]{Jenkins10b}. By choosing to minimize the CDPP in our light curves, we are optimizing them for planet searches. The CDPP is calculated for the ``CORR\_FLUX" light curve and is flattened using $\texttt{lightkurve.flatten()}$, which applies a 2\textsuperscript{nd} order Savitzky-Golay filter to the light curve. The data are binned on one hour timescales and the CDPP is evaluated using the built-in function \texttt{lightkurve.flatten().calculate\_cdpp()} using a window length of 51.  The filter is only applied to calculate the CDPP, and not used in the creation of the ``CORR\_FLUX" light curves saved in the \eleanor\ data product. The associated aperture and level of background subtraction for the minimum CDPP light curve are identified in the header of the \eleanor\ data product.

\subsubsection{Principal Component Analysis}
The \eleanor\ package also performs principal component analysis (PCA) to remove any additional systematics that are still potentially present and shared between nearby stars on the detector. PCA is a machine learning technique that calculates orthogonal eigenvectors between a set of input vectors, such as light curves. The Science Payload Operations Center (SPOC) pipeline \citep[]{jenkins16} performs calibration and systematics detrending for the short-cadence targets observed by \tess. In addition to producing TPFs and light curve files, the pipeline provides to the community co-trending basis vectors (CBVs), which represent systematic trends in the data. The CBVs are available through the Mikulski Archive for Space Telescopes (MAST). We take full advantage of the resources provided by the SPOC pipeline, such that we bin the two-minute CBVs into thirty-minute ones, in a similar fashion to how we create our quality flags.

The user has the ability to check the components themselves as well. The \eleanor\ light curve products include a ``PCA\_FLUX", which is created by subtracting the first 3 CBVs for the appropriate camera to minimize the possibility of astrophysical variability being imprinted in an eigenvector. All CBVs created by the SPOC pipeline are available for more artisanal analyses of light curves; the option to use more or fewer components is built into \texttt{eleanor.TargetData().pca()}. The PCA components are not stored in the \eleanor\ data products, but are stored on a forward-facing server that users have access to.

\subsubsection{Point-Spread Function Modeling}

We also include modeling of the point spread function of \tess\ as an option in \eleanor, stored as ``PSF\_FLUX.'' At present, this system models a uniform background level across the TPF, as well as an arbitrary number of Gaussians representing each star in the FOV. The relative positions of each Gaussian are set by the user, as the position of stars in the FOV is generally known to a very small fraction of a \tess\ pixel. The absolute position of the network of Gaussians is allowed to vary in each frame. We also fit a single width parameter in $x$ and $y$ as well as a single rotation angle. Each Gaussian has its own amplitude.
We then maximize the likelihood value of each parameter conditioned on the data in each frame using the \texttt{tensorflow} interface to the \texttt{scipy} implementation of the Truncated Newton minimizer, with options to maximize either a Gaussian or Poissonian likelihood function.

A single Gaussian does not, across most of the detector, accurately represent the shape of the actual PSF, which can have extended, asymmetric wings especially on the corners of the detector. However, sums of Gaussians centered on each star could be used to more accurately model the shape of the PSF. Regardless, even this simple but fast PSF model provides, in many cases, superior precision to aperture photometry methods, as can be seen for the case of WASP-100 in Figure \ref{fig:lightcurves}.
More sophisticated PSF models will be useful to accurately model stars in relatively crowded fields, and will be a focus point for future development of \eleanor.

Often, PSF modeling can provide higher precision light curves than just aperture photometry-based techniques, even for simple PSF models. This suggests there may be information in the terms that describe the shape of the PSF which can be used to improve our photometric precision.
As a proof of concept, we consider the WASP-100 system, as observed in Sector 1 of the \tess\ mission. 
At each cadence, we use the derived parameters of the PSF flux model that define the shape and rotation of the 2D Gaussian: its width in the row direction on the detector, its width in the column direction, and a cross term that defines the direction of the semi-major axis.
We then regress the ``RAW\_FLUX'' as calculated with \eleanor\ against each of those three time series in addition to the other vectors used in the standard determination of the ``CORR\_FLUX'' time series, using only the data when no quality flags are set.

\begin{figure*}[!ht]
\begin{center}
\includegraphics[width=1.0\textwidth,trim={0.0cm 0 0 0}]{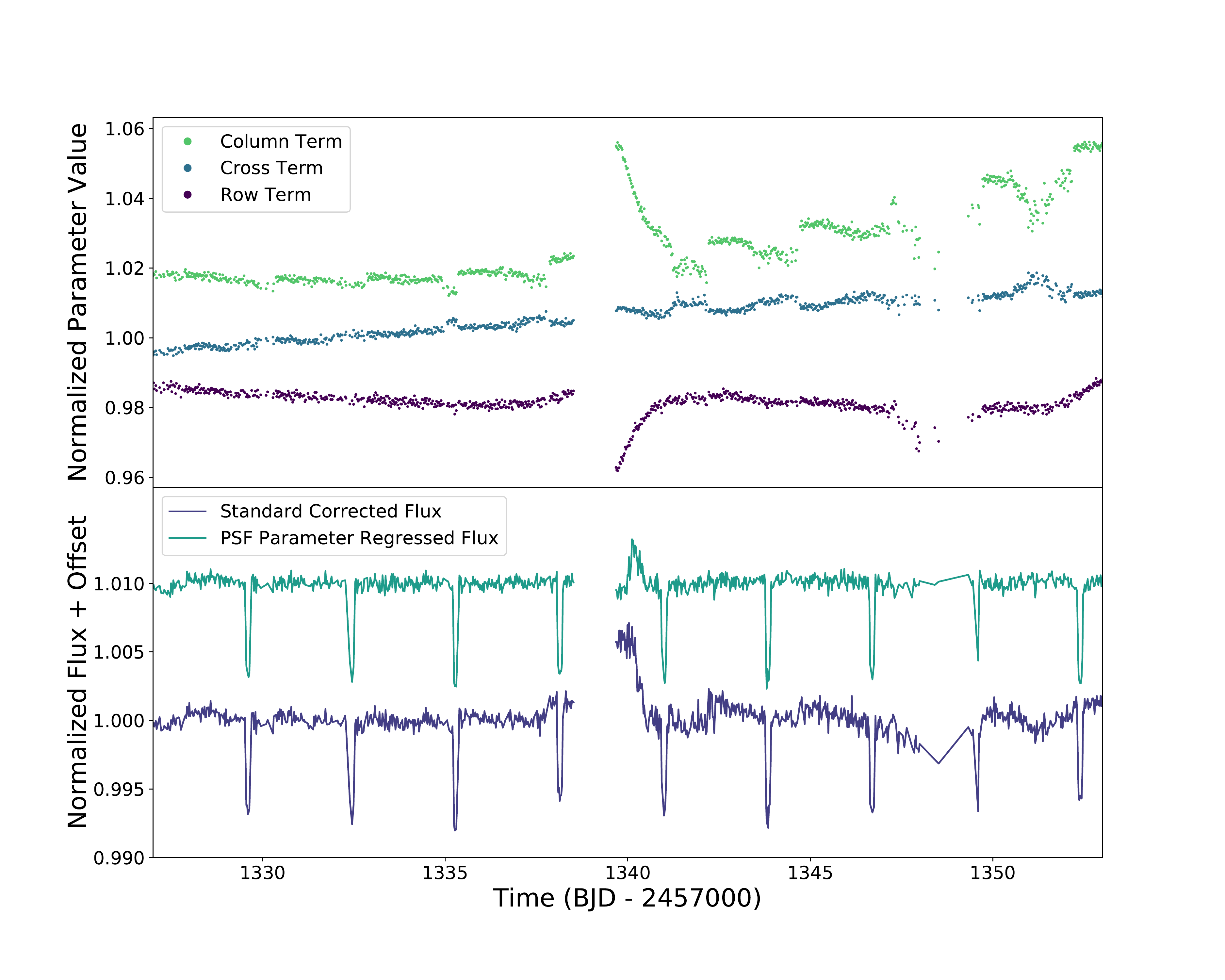}
\caption{(Top) Time series PSF parameters for the WASP-100 system, fitting a two-dimensional Gaussian to the pixel data for this star at each cadence. Long-timescale variations can be seen due to thermal variations, especially after data downlink. The periodic ``momentum dump'' thruster firings can also be clearly seen, especially in the size of the PSF along the column of the detector. (Bottom) Observed light curve for WASP-100, both using the standard regression against observed parameters and also including the PSF parameters in a linear model. The latter decreases the CDPP by more than 40\%, achieving nearly the same precision as fitting a PSF model to the data directly.}
\end{center}
\label{fig:PSFRegression}
\end{figure*}

The resultant time series photometry is shown and compared to the ``CORR\_FLUX'' in Figure \ref{fig:PSFRegression}. In this case, the PSF-regressed flux more effectively removes both long- and short-timescale effects. The standard corrected flux has a 1-hour CDPP of 279 ppm; while the PSF time series photometry has a CDPP of 161 ppm. The PSF-regressed flux has a CDPP of 162 ppm.
This method does not provide a significant improvement if a PSF model must \textit{a priori} be calculated, but potentially significant improvements in photometry for a large number of stars can be achieved by calculating the PSF parameters on a small number of targets and applying these to nearby stars, instead of calculating an explicit PSF model for each. 

The short-timescale changes in the shape of the stellar PSF may be due to pointing jitter on timescales shorter than the exposure time, causing an apparent expansion of the PSF when there is more jitter. This is bolstered by the fact that the apparent size of the PSF along the column direction appears to have sharp changes at the time of thruster firing events to dump momentum from the reaction wheels. 
In this case, future improvements may also be achieved by applying data from the high-cadence pointing information as recorded in quaternion form and provided as engineering data from the \tess\ spacecraft, as considered by \citep{Fausnaugh19}.

\subsection{The \eleanor\ Data Product}\label{subsec:products}

\begin{figure*}[!ht]
\begin{center}
\includegraphics[width=1.0\textwidth,trim={0.25cm 0 0 0}]{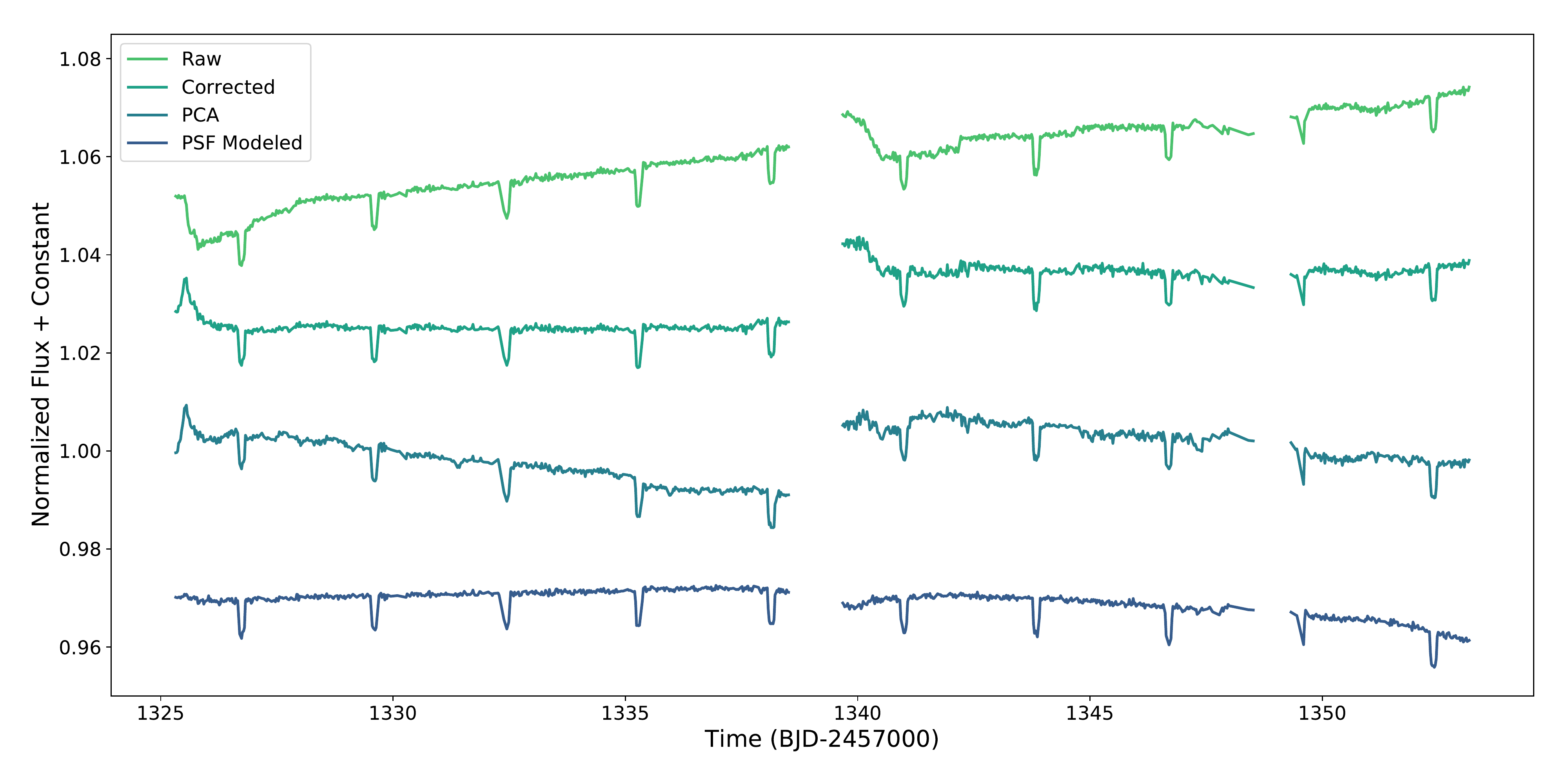}
\caption{An example of the four types of light curves that can be extracted and modeled using \eleanor. Raw flux is the sum of pixels in the aperture. Corrected flux is raw flux with a linear regression as a function of pixel location, background, and time. PCA flux is a Principal Component Analysis subtracted flux, to remove common systematics between targets on the same camera. PSF modeled flux is the 2D Gaussian point-spread function modeled flux. The 10 transits for WASP-100 are clearly seen in all four light curves. The gaps in the light curves result from two different sources: The first gap is the result of data down-link and re-pointing to the same position on the sky. The second gap is the result of the telescope losing fine-pointing. The second gap has been masked by our quality flags. \label{fig:lightcurves}}
\end{center}
\end{figure*}

The final data product is stored as a Flexible Image Transport System (FITS) file. Each FITS file contains the cadence-stacked, background-subtracted flux pixels for a 13 $\times$ 13 region, centered on the source, and the cadence-stacked flux error pixels for the same region. The files contain all 21 aperture masks tested in the light curve extraction process as well as the raw and corrected fluxes for these apertures. For the automatically selected best aperture, three light curves are available: ``raw'' flux, ``corrected'' flux regressed against instrumental effects, and ``PCA'' flux with common modes subtracted. Users also have the option to create a PSF modeled light curve with the \eleanor\ software package. However, the PSF flux is not a default light curve in the data product due to the relatively large processing time required. An example of each type of light curve can be seen in Figure \ref{fig:lightcurves}, which additionally shows the recovery of the known planet WASP-100b \citep[]{hellier14, Stassun17}.

In addition to photometric information, the produced files contain the $x$ and $y$ centroid positions, as inferred by our pointing model, and quality flags, based on the two-minute cadence targets and our own quality flag for cadences, as discussed previously. We create light curve files for sources in the TIC with I $\leq$ 16. This includes potentially saturated stars, which should be handled cautiously by the user. See Section~\ref{subsec:limits} for more information on \eleanor\ limitations. Members of the community can use the \eleanor\ package to create light curves for fainter or extragalactic objects, or for a more detailed or optimized analysis of individual objects.

\section{Results}\label{sec:results}

\begin{figure*}[!ht]
\begin{center}
\includegraphics[width=1.0\textwidth,trim={0.25cm 0 0 0}]{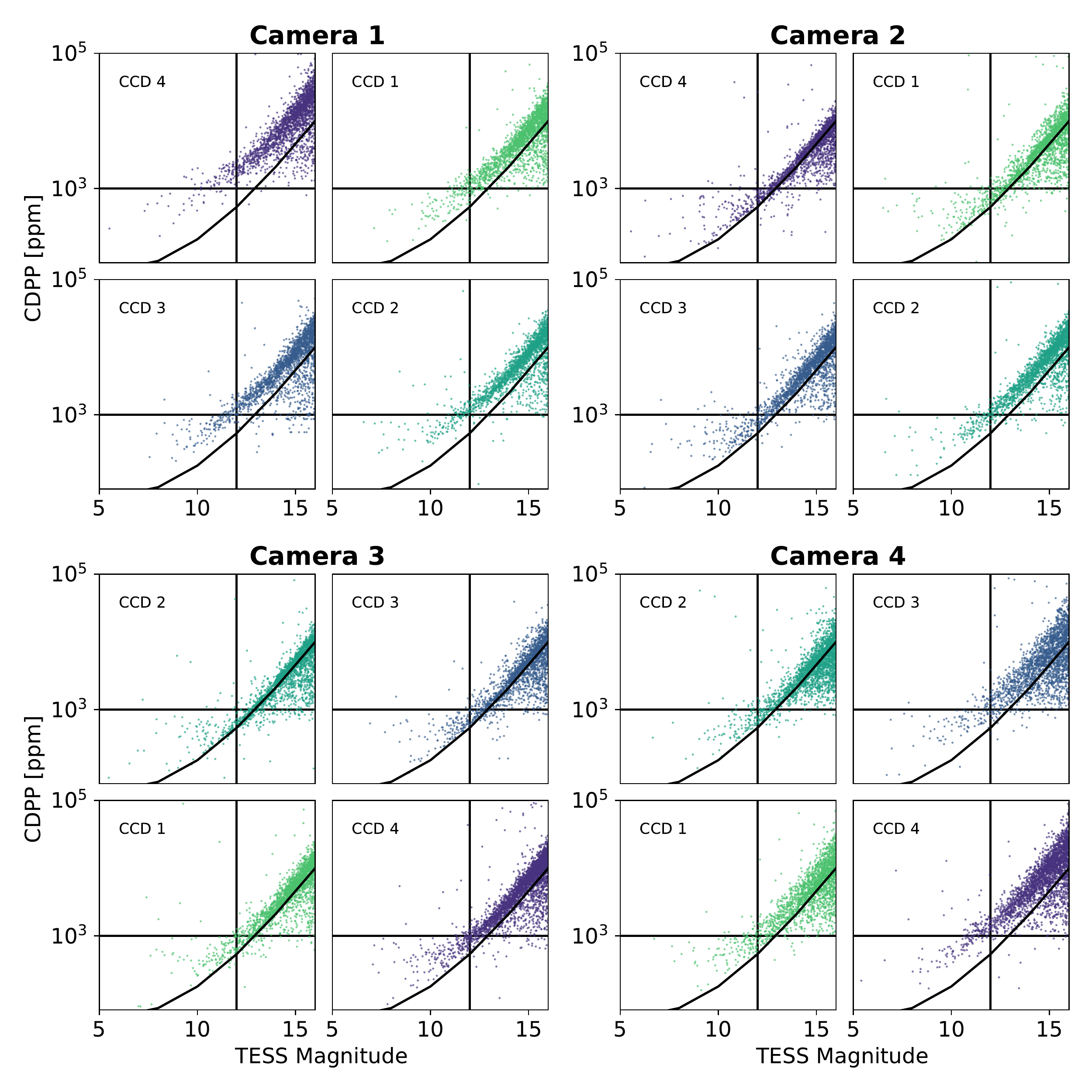}
\caption{The CDPP as a function of \tess\ Magnitude for 32,000 stars on each camera. Stars are colored by which CCD they fall on. Camera 4 (see Figure \ref{fig:ffi}) experiences significantly more background features than the other three cameras, leading to an overall increase in CDPP. The colors correspond to the CCD. We provide guiding dashed lines at T\textsubscript{mag} = 12 and CDPP = 10\textsuperscript{3}\,ppm to allow for easier comparison between CCDs. CCDs with less background have lower CDPP values than those with more prominant background features. We additionally include the total predicted noise level from \cite{sullivan15} as the solid black line across all plots. \label{fig:mag_cdpp}}
\end{center}
\end{figure*}

We calculated the 1-hour CDPP for 32,000 light curves in each of \tess's four cameras in Sector 1. The CDPP as a function of \tess\ magnitude is shown in Figure \ref{fig:mag_cdpp}. We subdivide into CCDs to demonstrate how light curves extracted from regions with more significant background effects (see Camera 4, CCD 1 in Figure~\ref{fig:ffi}) are affected.

The CDPP remains fairly consistent for all CCDs in a given camera, with the exception of Camera 4. CCD 4 experiences noticeably more systematics than the other CCDs, leading to an overall increase in CDPP values. Note that the presence of the Large Magellanic Cloud in CCDs 1 and 2 does not lead to a significant difference in CDPP values when compared to other cameras; while diffuse light from the LMC is obvious, it is stable.

We further investigated the faint, $T_{mag}\,>$\,12 stars that fall below the general trend in Figure~\ref{fig:mag_cdpp}. Across all 4 cameras, we found 2,200 fell within this range. Of this sub-sample, 90\% have brighter neighboring stars within 50", or roughly two \tess\ pixels. It can therefore be concluded that the nearby star is contaminating the aperture being used for these fainter stars and therefore decreasing their overall CDPP. 

\subsection{Comparison to Other Pipelines}

\begin{figure*}[!ht]
\begin{center}
\includegraphics[width=1.0\textwidth,trim={0.25cm 0 0 0}]{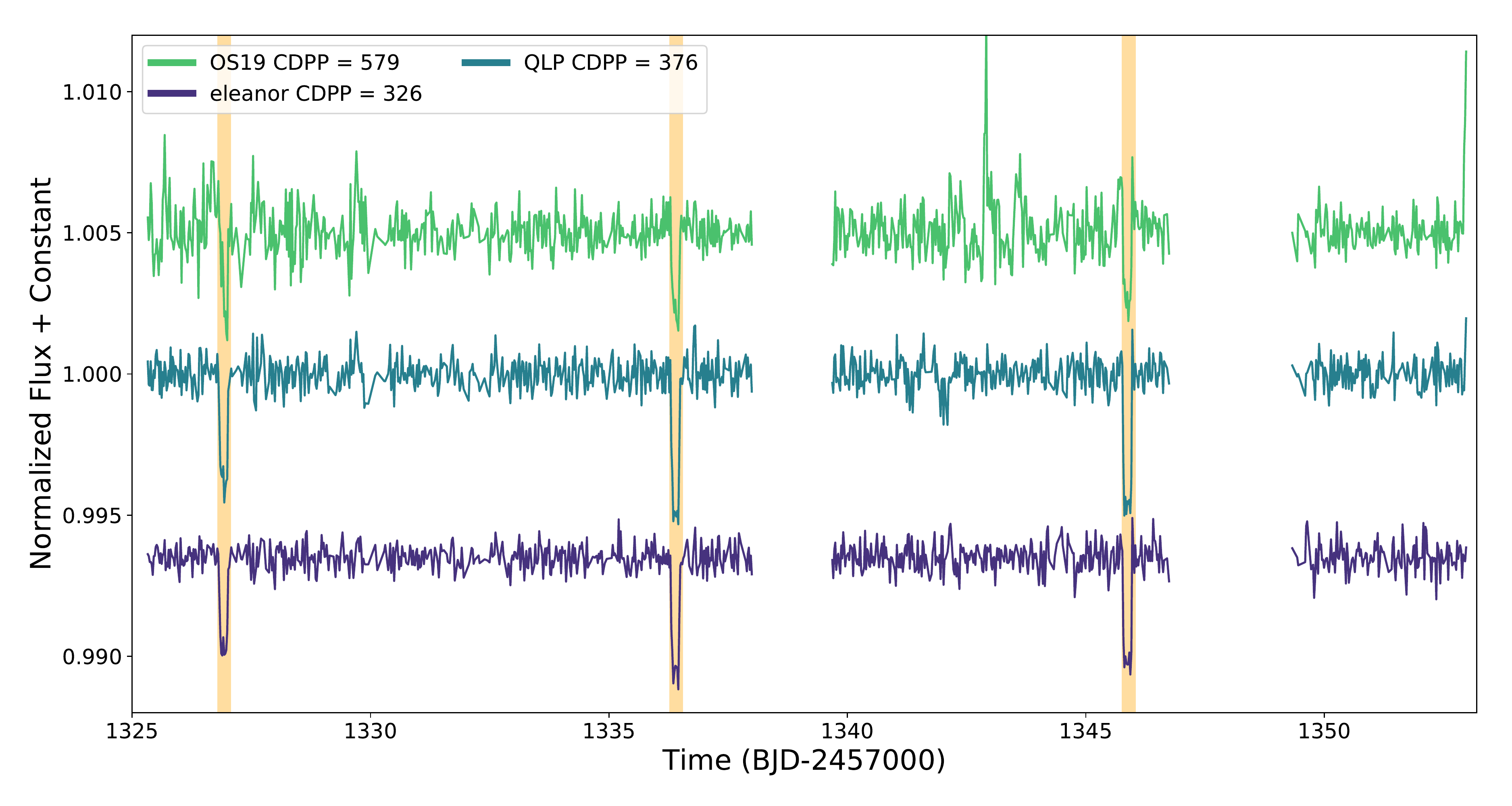}
\caption{A comparison of results from using three separate tools to extract light curves from the \tess\ FFIs: \citet[][OS19; green]{oelkers19}, MIT's Quick-Look Pipeline (QLP; blue), and \eleanor\ (purple). Each light curve received the same quality flags and flattening treatment when calculating the CDPP. The transits were removed from the calculation and are highlighted in the figure to demonstrate that all three pipelines recover it. TIC29857954 is a T$_{mag}$ = 10.71 host star. The \eleanor\ light curve has the lowest CDPP of the three pipelines. \label{fig:pipelines}}
\end{center}
\end{figure*}

We compare the \eleanor\ light curve CDPP to that of \cite[][OS19]{oelkers19} and the MIT Quick-Look Pipeline (QLP) to assess its performance in removing instrumental and astrophysics systematics. The QLP light curves are not available to the public except for a small number of published, confirmed planets; for comparison we use the published light curve of TOI-172\,b \citep{rodriguez19}.  We apply the \eleanor\ quality flags across all light curves for a uniform comparison. We apply additional masks for times when the observatory went out of its fine-pointing mode (1338 $<$ Time $<$ 1339 and 1346.8 $<$ Time $<$ 1348.6) and for the 3 transits of TOI-172 b, marked in Figure \ref{fig:pipelines} by the vertical orange lines, to quantify the level of scatter in the light curve out of transit. We show the transits in Figure \ref{fig:pipelines} to demonstrate \eleanor's transit detection capabilities. Each light curve has been flattened using identical processes.

For TOI-172, a $T_{mag} = 10.71$ host star, the \cite{oelkers19} light curve has CDPP\,=\,579~ppm; the 
QLP light curve has CDPP\,=\,376~ppm; and the \eleanor\ light curve has CDPP\,=\,325~ppm. The smaller scatter in the \eleanor\ light curves will enhance the community's ability to detect small planet transit signals in the FFIs.

\begin{figure*}[!ht]
\begin{center}
\includegraphics[width=0.5\textwidth,trim={0.25cm 0 0 0}]{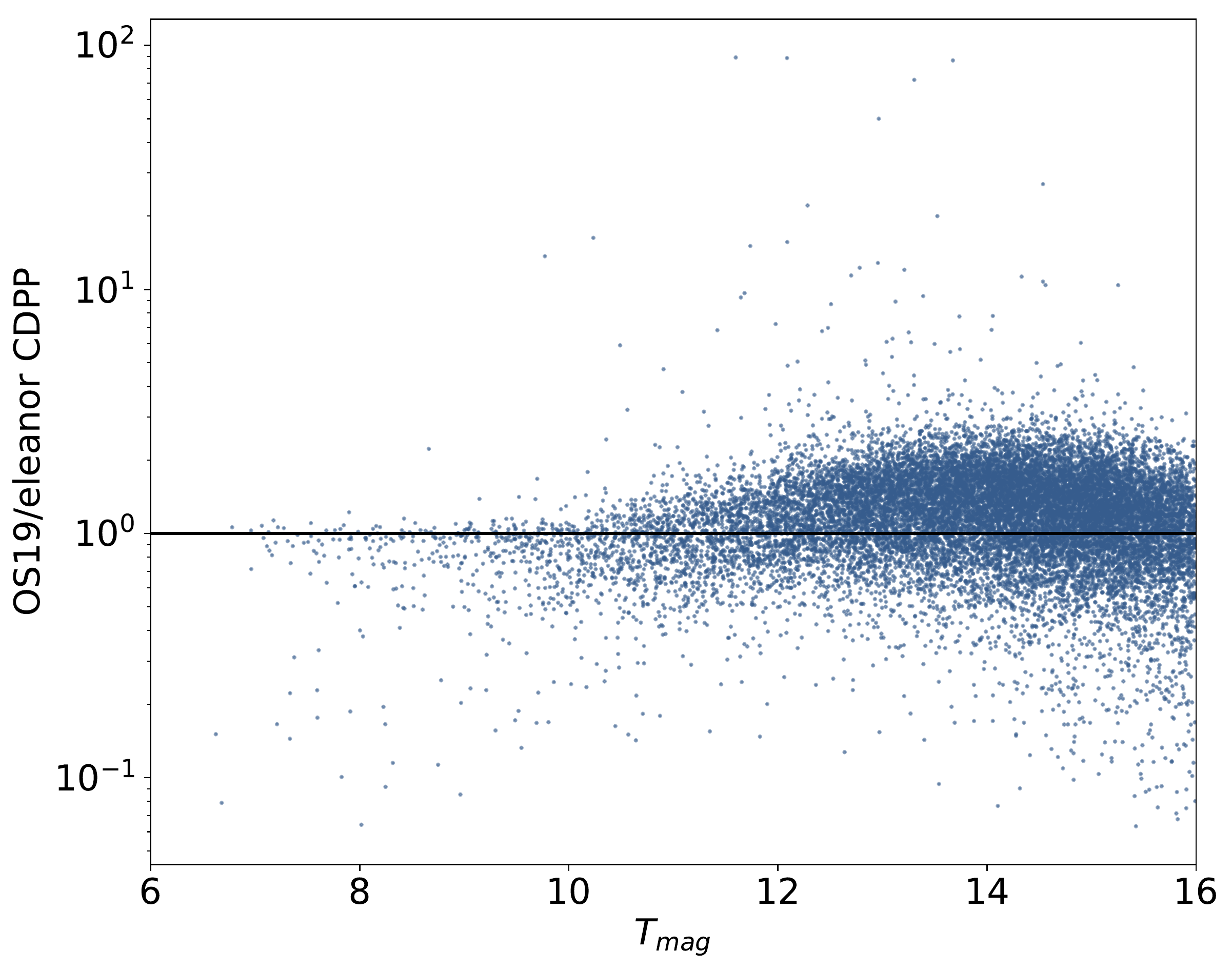}
\caption{We calculated the 1-hour CDPP for 17,000 of the same stars between the \eleanor\ and OS19 light curves. Approximately 12,000 of the 17,000 stars compared in this figure have OS19/\eleanor\ CDPP $>$ 1. This is most noticeable in the $12 \leq T_{mag} \leq 14$ regime. Points which fall along the one-to-one line (black) are being handled equally well by both pipelines. \label{fig:pipeline_cdpp}}
\end{center}
\end{figure*}

We complete a more quantitative comparison of our light curves to that of OS19. For Sector 1, OS19 has created light curves for 1.2 million stars. We cross-matched the light curves we created for Figure~\ref{fig:mag_cdpp} and found 17,000 stars of the same TIC ID. We calculated the CDPP of both light curves in the same way as described above. The results can be seen in Figure~\ref{fig:pipeline_cdpp}. Overall, approximately 12,000 of the stars compared have OS19/\eleanor\ CDPP $>$ 1, suggesting the \eleanor\ light curves have less scatter. This is especially true for $12 \leq T_{mag} \leq 14$ regime, as seen in the figure.

\subsection{Recovery of Known Planets}

\begin{figure*}[!ht]
\begin{center}
\includegraphics[width=1.0\textwidth,trim={0.25cm 0 0 0}]{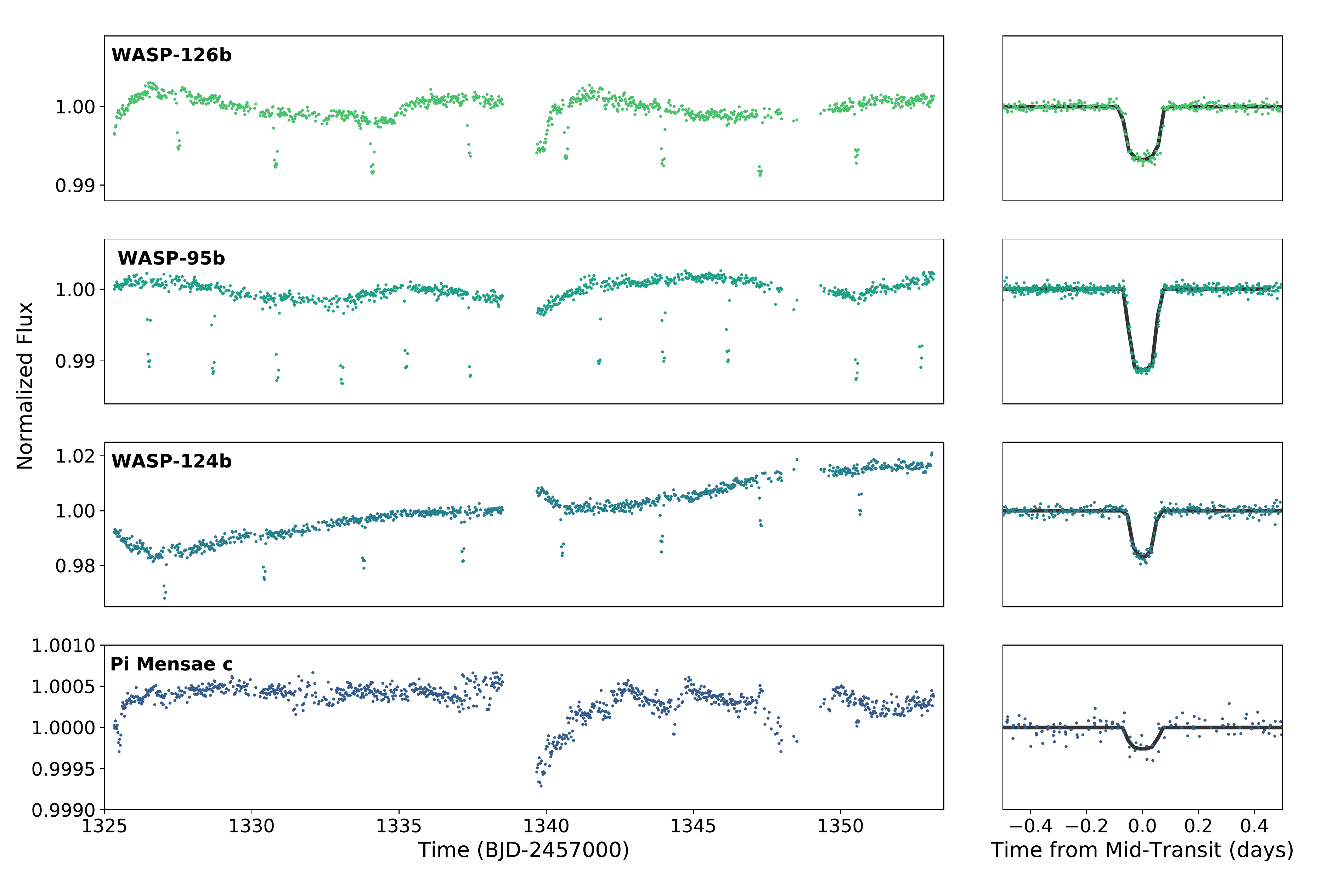}
\caption{An example of the recovery of four known planets in the FFIs. We plot the normalized \eleanor\ PSF-modeled flux with our quality flags. Transit fits modeled with \texttt{batman} are shown by the solid black lines in the phase-folded subplots. Retrieved system parameters are quoted in Table~\ref{tab:planetparameters}. The data gaps correspond to those described in Figure~\ref{fig:lightcurves} \label{fig:wasps}}
\end{center}
\end{figure*}

In order to further demonstrate the quality of the \eleanor\ light curves, we recover a few known transiting planets that were observed in Sector 1. We show the light curves for four of these planets in Figure~\ref{fig:wasps}. We use the \texttt{batman} \citep[]{kreidberg15} transit fitting software package, following the methods from \cite{Mandel02}, to derive planet parameters from the \eleanor\ light curves and compare our results with the known parameters. The planet parameters we derived are consistent with values quoted in \cite{maxted2016} and \cite{Stassun17} for the WASP planets, and with \cite{Huang18} and \citet{gandolfi18} for $\pi$ Mensae c and are quoted in Table~\ref{tab:planetparameters}.

We complete a more extensive Markov Chain Monte Carlo (MCMC) analysis for WASP-126b \citep[]{maxted2016} using \texttt{emcee} \citep[]{Foreman-Mackey12}, an implementation of the affine-invariant ensemble sampler of \cite{Goodman10}. We initialized our MCMC run with the best-fit parameters from a single \texttt{batman} fit and ran it for 200 steps, to complete a burn-in. The parameters from the 200\textsuperscript{th} run were then used as the starting point for the next run of 1500 steps. The parameters and uncertainties are quoted in Table \ref{tab:planetparameters}. All system parameters derived with the \eleanor\ light curve fall within 1$\sigma$ of the accepted parameters from \cite{maxted2016}, with the exception of $a/R_*$, which agrees within 1.5$\sigma$. Overall, the derived parameters for WASP-126 correspond well to parameters in the literature.

\subsection{New Science with \textbf{eleanor} Light Curves}

We demonstrate the potential for new science to come out of the \tess\ FFIs using \eleanor\ by performing a limited search for periodic signals in Sector 1 data. We performed a search of 12,000 stars using the box-least squares (BLS; \texttt{bls.py}\footnote{https://github.com/dfm/bls.py}) module in \texttt{astropy} to phase fold light curves and identify new periodic candidates, ranging from new planet candidates to RR Lyrae stars. We searched a period range of 0.5 to 8 days and recorded the candidates where the maximum peak in the periodogram (period vs. log likelihood) was more than 9$\sigma$ above the mean. 

\begin{figure*}[!ht]
\begin{center}
\includegraphics[width=1.0\textwidth,trim={0.25cm 0 0 0}]{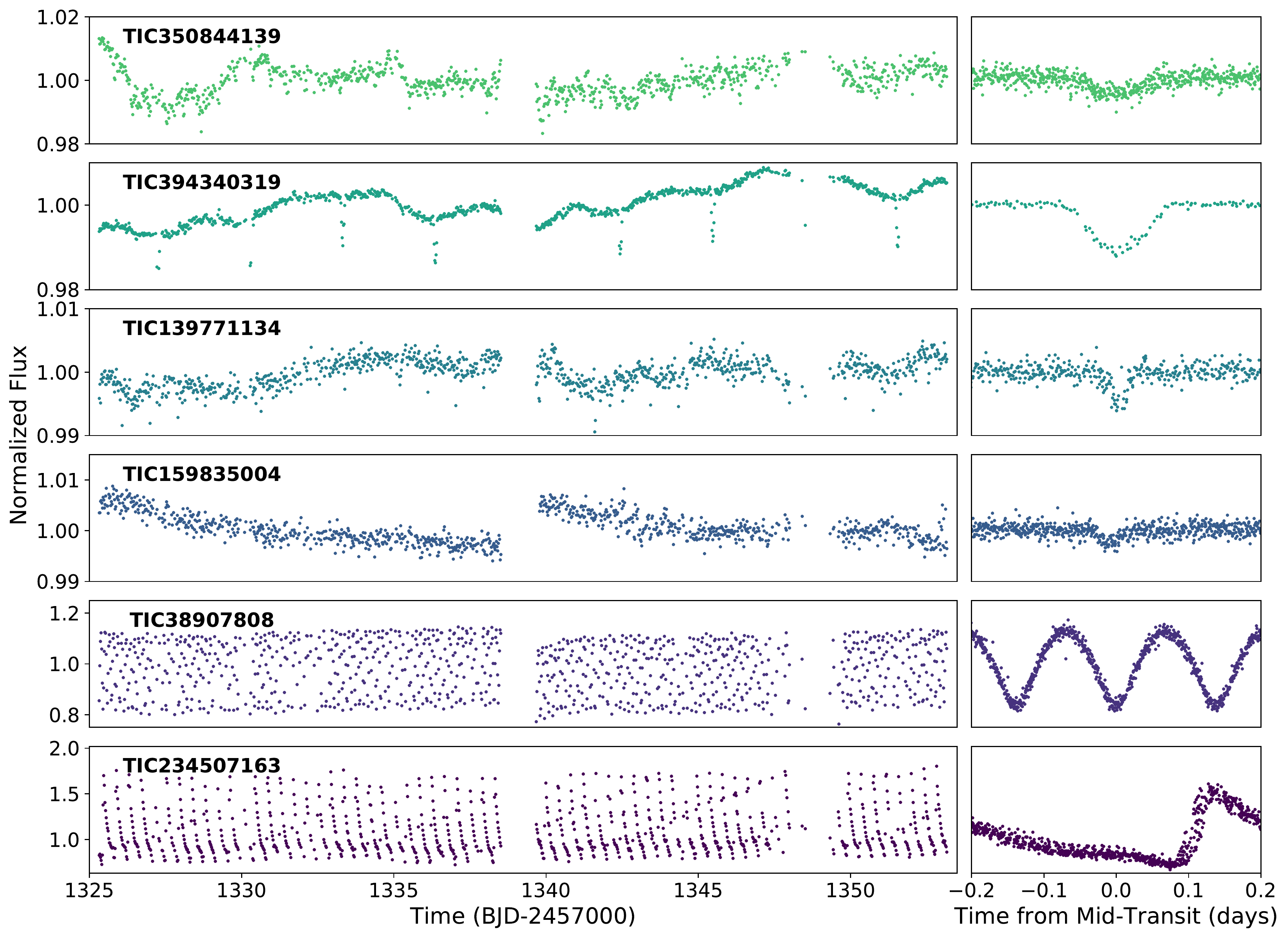}
\caption{Six examples of new periodic astrophysical signals identified in the \tess\ FFIs using \eleanor. The first four rows represent previously unidentified planet candidates; the fifth row represents a new contact binary; the sixth row represents a previously known RR Lyrae. We completed a simple BLS fit to each light curve to obtain the time from transit center, period, and depth, which are presented in Table \ref{tab:newplanets}. This population represents a small subset of sources, which will all be published in later work and available for the community to use. \label{fig:new_stuff}}
\end{center}
\end{figure*}

The identified candidates with periodic signals were then vetted by-eye. The candidates that passed this check can be found in Table \ref{tab:newplanets}. We present the properties from the BLS fits and uncertainties for new planet candidates and eclipsing binaries. This is an incomplete list of new exoplanet candidates identified with \eleanor. A full list of candidate signals will be included in future work. An example of the light curves with candidate periodic signals can be seen in Figure \ref{fig:new_stuff}.

\begin{figure*}[!ht]
\begin{center}
\includegraphics[width=1.0\textwidth,trim={0.25cm 0 0 0}]{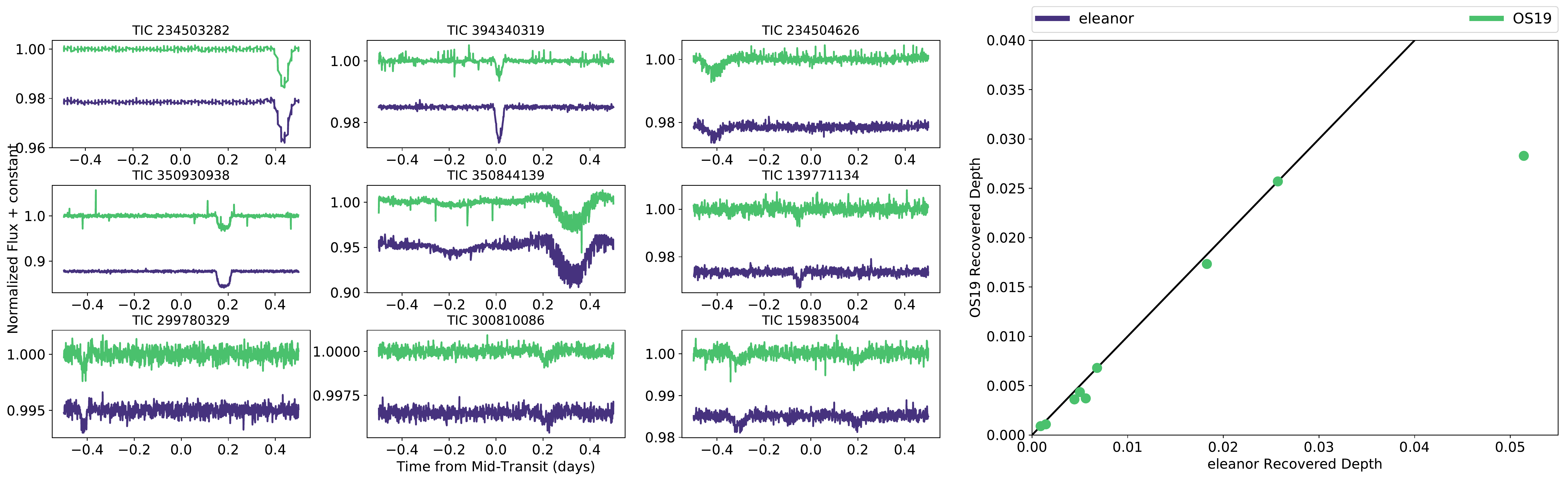}
\caption{We completed a search of the new planet candidates identified in Table~\ref{tab:newplanets} within the OS19 light curves. All light curves show transits folded on the same period. The right panel is a comparison of the recovered planet transit depths using \texttt{batman} fits. Most depths are in agreement, tending towards slightly deeper in the \eleanor\ light curves. However there is a significant difference in that of TIC 350844139, which may be due to dilution in the extracted OS19 light curve from a nearby star. Green light curves are from OS19; purple light curves are from \eleanor. \label{fig:planet_depth_comp}}
\end{center}
\end{figure*}

We searched for the newly identified planet candidates in the OS19 light curves. All light curves except 2, TIC 349155660 and TIC 349832804, were available. We compare the light curves and recover transit depths using \texttt{batman} transit fits for both \eleanor\ and OS19 in Figure~\ref{fig:planet_depth_comp}. All depths are in good agreement, with a tendency towards a deeper transit using \eleanor, except for that of TIC 350844139. This discrepancy could be the result of dilution in the OS19 light curve from a nearby star.

We note that \cite{sullivan15} concluded that there will be roughly 1000 false positives within the 2-minute targets. \cite{Barclay18} conducted an additional analysis of the false positive rate for the FFIs and concluded the false positive rate should increase from 1 false positive per 180 stars in the 2-minute targets to nearly 5 times that for the FFIs, but could increase to nearly 11 times that, depending on the parameter space probed for planet transits. Therefore, further vetting is required for the candidates identified in this work. All planet candidates identified with \eleanor\ will be hosted on ExoFOP-TESS as Community \tess\ Objects of Interest (CTOIs).

\begin{figure*}[!ht]
\begin{center}
\includegraphics[width=1.0\textwidth,trim={0.25cm 0 0 0}]{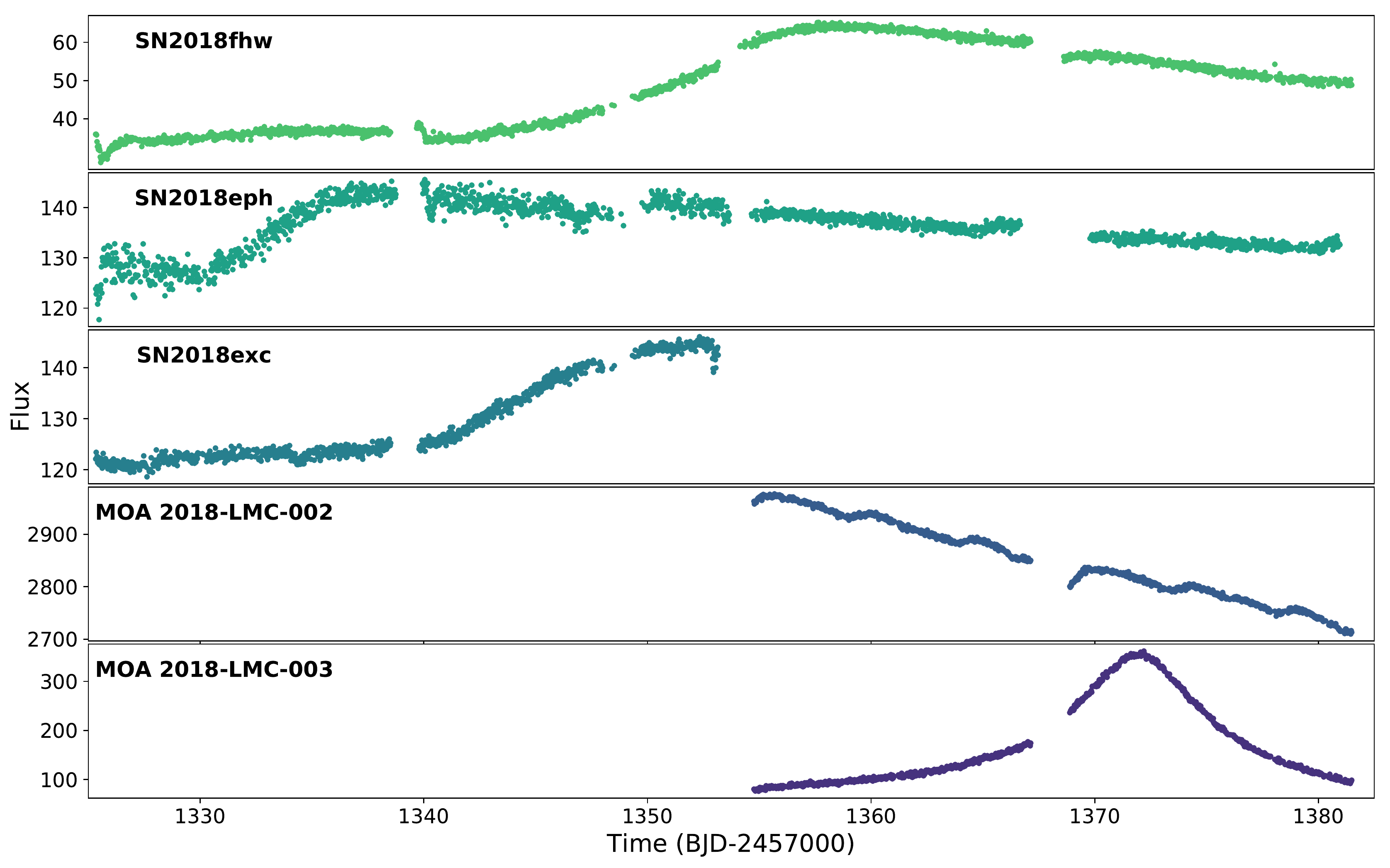}
\caption{Example of five recovered transient events in \tess\ Sectors 1 and 2. SN2018fhw and SN2018exc are known Type Ia supernovae; SN2018eph is a known Type II supernova; MOA 2018-LMC-003 is a previously identified microlensing event within the LMC; the event causing the light curve for MOA 2018-LMC-002 is unknown. Even though the \eleanor\ light curves are tailored towards finding new transiting exoplanets, the software can be used for other astrophysical detections. \label{fig:supernovae}}
\end{center}
\end{figure*}

In addition, we demonstrate the use of \eleanor\ to explore extragalactic astrophysics by recovering known supernovae that occurred in Sectors 1 and 2 (Figure \ref{fig:supernovae}). SN2018fhw \citep[]{brimacombe18, eweis18} and SN2018exc \citep[]{stein18, tonry18} are classified as Type Ia supernovae; SN2018eph \citep[]{brimacombe18-2, onori18} is classified as a Type II supernova; MOA 2018-LMC-002 is an unknown astrophysical event; MOA 2018-LMC-003 is a known microlensing event towards the Large Magellanic Cloud. By having light curve information before the triggering of the supernova event, the photometric information from \tess\ can be used to infer information about the supernova progenitor. This figure demonstrates how one can use the \eleanor\ software to recover extragalactic events. However, for a better understanding of how to obtain more accurate supernova light curves, see \cite{Fausnaugh19}. Furthermore, uniform high-cadence light curves will be useful to search microlensing events for the signatures of planetary companions to the lensing stars. 

\subsection{Software Limitations}\label{subsec:limits}

\begin{figure*}[!ht]
\begin{center}
\includegraphics[width=0.7\textwidth,trim={0.25cm 0 0 0}]{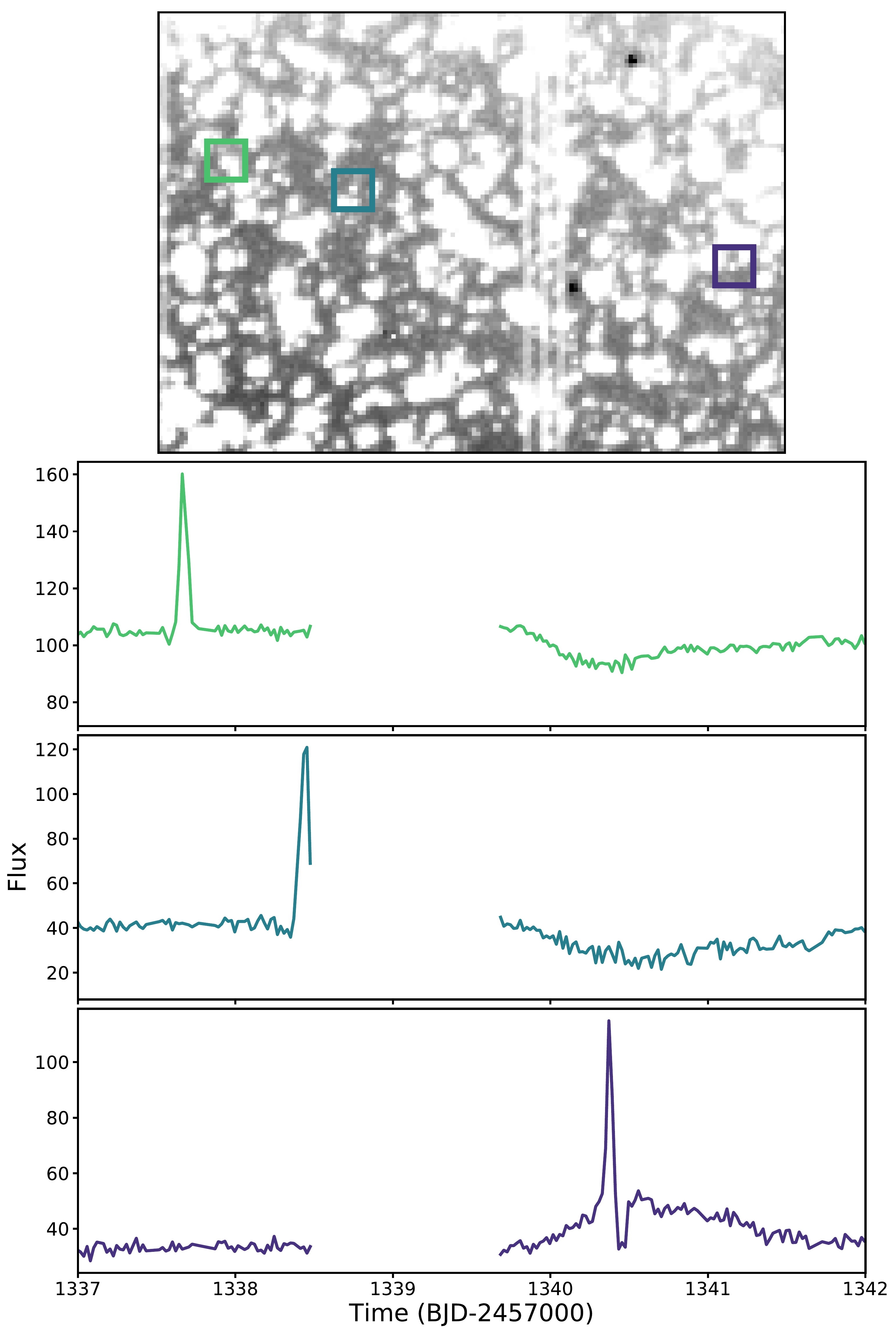}
\caption{An example of an asteroid crossing through a light curve for three stars on the same postcard. There is a clear increase in flux as the asteroid passes through the aperture. The TPF and associated light curves are identified by the same color. The crossing occurs before and after the data downlink gap between the two orbits during Sector 1. \label{fig:asteroid}}
\end{center}
\end{figure*}

There are several cases in which \eleanor\ light curves may produce non-optimal results. The first is the presence of moving solar system objects. We do not perform any corrections for when Mars saturated a significant fraction of one detector in Sector 1 or when asteroids pass through an aperture for a given target (see Figure \ref{fig:asteroid} as an example). An asteroid induces a clear spike in the raw light curve and its location can be traced as it passes in front of different stars on this postcard as a function of time. For stars lying in or near the ecliptic plane, we recommend spikes in any light curve be checked to ensure they are not a foreground Solar System object.

The saturated star \citep[$T \leq 6.8$,][]{sullivan15} on the detector are beyond the intended scope of \eleanor. Although we create \eleanor\ data products for these targets, we recommend the user look to see if the target has been observed at 2-minute cadence (which is likely), create a custom aperture larger than the apertures in the \eleanor\ default library (Fig. \ref{fig:aps}), or use a halo aperture approach \citep[]{white17}. Additionally, PSF modeling of saturated stars will produce poor results as the PSF model is not an accurate representation of the behavior of the star on the detector in these cases. Additionally, we predict that roughly $\sim$6\% of faint ($T_{mag} > 14$) stars observed every sector will have significant contamination from nearby bright neighboring stars, leading to a lower CDPP that falls below the predicted noise limit (see Fig.~\ref{fig:mag_cdpp}).

In the case of crowded fields, we limit the default apertures tested to those with pixel sums $<$ 9 (Apertures A-J, see Figure~\ref{fig:aps}) to mitigate the possibility of a nearby star contamination the aperture too significantly. The crowded field flag can be manually set by a user, or the user can restrict themselves to only considering the apertures used in the crowded field setting in their own analysis of the \eleanor\ data products. For the light curves considered in this work, we explicitly set the crowded field flag if and only if the \tess\ magnitude of the star is fainter than $T$ = 14.0. In all cases, it is recommended to investigate the aperture and the locations of other nearby stars to ensure the reliability of the light curve. For additional information on other spacecraft issues and warnings that may affect an extracted \eleanor\ light curve, see the \tess\ Science Data Products Description \citep[]{tenenbaum2018}.

\section{Data Availability and Software Tools}\label{sec:availability}

For each sector in the Southern Hemisphere, we will create and release \eleanor\ data products (See \ref{subsec:products}) to the community. After TESS completes all observations in the southern hemisphere, we will reprocess all light curves for a uniform library. This also allows us to potentially improve earlier light curves as we continue to learn about new methods to remove the background, determine the pointing of the spacecraft, and manage crowded regions of the detector. The data will be available as a high level science product (HLSP) at the Mikulski Archive for Space Telescopes (MAST).

In addition to our light curve products, \eleanor\ is an open-source package that can be downloaded through GitHub or via the Python Package Index. \eleanor\ was designed with the user in mind. Users have the ability to create TPFs and light curves for any source or position on the detector by passing in a TIC ID, Gaia ID, or RA and Dec coordinates. When creating a new TPF, \eleanor\ will go through all of the steps described above for the user’s request.

Beyond the standard steps described in Section 2, there are several features the user can customize if they make a light curve with the \eleanor\ software. Users can set the size (height $\times$ width) of the TPF \eleanor\ extracts. The height and width are forced to be odd numbers to allow the target to be at or very near the center of the cutout. Users can call \texttt{do\_psf = True} when initializing the light curve object, which allows for a PSF modeled light curve in the output. Additionally, the user has the ability to set the region around a source that they want to base the background subtraction on. The default background size is the size of the TPF. For the publicly available light curve products, this will be $13 \times 13$ pixels (see Appendix~\ref{appendix:code} for syntax examples in Python using \eleanor). 

After each sector is observed, we will search for new planet candidates within the FFIs. Light curves that have been identified as planet candidates will be hosted on ExoFOP-TESS. We will release a catalog of new planet candidates as well as other interesting astrophysical events, such as eclipsing binaries and RR Lyraes. This information will be open to the community. Although we are predominantly interested in finding new planet candidates, our hope is these light curves will yield scientific discoveries across many branches of astrophysics, including supernovae characterization and identification of stellar oscillations, to name a few. The open-source \eleanor\ products will be available for a diverse set of scientific discoveries that will be achievable using the \tess\ FFIs. 

\acknowledgements

We thank Doug Caldwell, Michael Fausnaugh, Jon Jenkins, and Roland Vanderspek for valuable discussions. We thank Patrick Vallely for his direction to recovering known supernovae in the \tess\ field of view. Work by B.T.M. was performed under contract with the Jet Propulsion Laboratory (JPL) funded by NASA through the Sagan Fellowship Program executed by the NASA Exoplanet Science Institute. We would like to acknowledge the live animal webcams of the Kansas City Zoo, and especially the emotional support of Elvis the emperor penguin during our hack weeks.
This work was funded in part through the NASA \tess\ Guest Investigator Program, as a part of Program G011237 (PI Montet).

This paper includes data collected by the TESS mission, which are publicly available from the Mikulski Archive for Space Telescopes (MAST).
Funding for the TESS mission is provided by NASA’s Science Mission directorate.

This project was developed in part at the \textit{Building Early Science with} TESS meeting, which took place in 2019 March at the University of Chicago.

\software{%
    numpy \citep{numpy},
    matplotlib \citep{matplotlib},
    scipy \citep{jones01}
    tensorflow \citep{tensorflow},
    lightkurve\footnote{https://doi.org/10.5281/zenodo.2557026},
    photutils\footnote{https://doi.org/10.5281/zenodo.2533376},
    astropy\citep{astropy:2013, astropy18},
    eleanor\footnote{https://doi.org/10.5281/zenodo.2597620},
    sklearn\citep{sklearn11}
    }

\facility{TESS}

\appendix 
\section{\eleanor\ Software Demonstration} \label{appendix:code}

This represents an example of the basic commands for \eleanor\ light curves for one target given a known TIC ID. The user has the ability to set the sector they wish to create a light curve for, if the target has been observed in multiple sectors.

\begin{lstlisting}[language=Python]
# Import the eleanor package
import eleanor

star = eleanor.Source(tic=38846515, sector=1)

# PSF Modeling is a function that needs to explicitly be set to True
# Default is no PSF modeled flux
data = eleanor.TargetData(star, do_pca=True, do_psf=True)

# Calling eleanor quality flags
q = data.quality == 0

# Calling the different flux options
raw_flux  = data.raw_flux[q]
corr_flux = data.corr_flux[q]
pca_flux  = data.pca_flux[q]
psf_flux  = data.psf_flux[q]

# Light curve with the minimum CDPP
best_aperture = data.aperture

# Saving your TPF file and light curves
data.save()
\end{lstlisting}

Users can additionally create \eleanor\ light curves based on a given set of coordinates that have been observed in multiple sectors.

\begin{lstlisting}[language=Python]
# Calling a light curve for a set of coordinates
from astropy.coordinates import SkyCoord, Angle
from astropy import units as u

ra = Angle(68.959732, u.deg)
dec = Angle(-64.02704 , u.deg)

# Obtaining target information for multiple sectors
# Returns a list of eleanor.Source() objects
stars = eleanor.multi_sectors(coords=SkyCoord(ra=ra, dec=dec), sectors=[1,2])

# Extracting light curves
data0 = eleanor.TargetData(stars[0])
data1 = eleanor.TargetData(stars[1])

# Creating a custom circular aperture of radius 1
eleanor.TargetData.custom_aperture(data0, shape='circle', r=1)

# Recreates default light curve with new aperture
eleanor.TargetData.get_lightcurve(data0)

# Creating a PCA Model; sets data0.psf_flux
eleanor.TargetData.pca(data, flux=corr_flux, modes=4)

# Creating a PSF Model; sets data1.psf_flux
eleanor.TargetData.psf_lightcurve(data1, model='gaussian', likelihood='poisson')

# Available light curves
# No PSF modeled flux was created for this eleanor object
raw_flux0 = data0.raw_flux # with new aperture
corr_flux0 = data0.corr_flux # with new aperture
pca_flux0 = data0.pca_flux

# No PCA corrected flux was created for this eleanor object
raw_flux1 = data1.raw_flux
corr_flux1 = data1.corr_flux
psf_flux1 = data1.psf_flux


\end{lstlisting}


\begin{deluxetable}{l r r r r r r}[!ht]
\tabletypesize{\footnotesize}
\tablecaption{Transit Properties of Recovered Planets \label{tab:planetparameters}}
\tablehead{
\colhead{Planet} & \colhead{$R_p/R_*$} & \colhead{P} & \colhead{$a/R_*$} & \colhead{i} & \colhead{e} & \colhead{Reference}
}
\startdata
WASP-126b & 0.0776 & 3.28880 & 7.634 & 87.9 & $<$0.18 & \cite{maxted2016}\\
WASP-126b & 0.076$^{+0.003}_{-0.002}$ & 3.28692$^{+4e-5}_{-2e-5}$ & 6.080$^{+1.162}_{-1.307}$ & 86.972$^{+2.117}_{-3.943}$ & 0.183$^{+0.263}_{-0.135}$ & This work\\
\hline
WASP-95b & 0.1024 & 2.1847 & 6.51 & 88.4 & 0.0 & \cite{Stassun17}\\
WASP-95b & 0.099$^{+0.002}_{-0.002}$ & 2.18471$^{+6e-5}_{-6e-5}$ & 6.0996$^{+0.3387}_{-0.4278}$ & 87.627$_{-2.420}^{+1.687}$ & 0.047$_{-0.033}^{+0.037}$ & This work\\
\hline
WASP-124b & 0.12   & 3.3727 & 9.434 & 86.3 & $<$0.017 & \cite{maxted2016}\\
WASP-124b & 0.119$^{+0.006}_{-0.006}$   & 3.3729$^{+0.0009}_{-0.0008}$ & 9.420$^{+1.391}_{-1.151}$ & 86.546$^{+2.247}_{-1.376}$ & 0.094$^{+0.070}_{-0.063}$ & This work\\
\hline
$\pi$ Men c & 0.017 & 6.26790 & 13.38 & 87.456 & 0.0 & \cite{Huang18}\\
$\pi$ Men c & 0.017 & 6.26834 & 13.10 & 87.31  & 0.0 & \cite{gandolfi18}\\
$\pi$ Men c & 0.0182$_{-0.0001}^{+0.0001}$ & 6.2720$_{-0.0001}^{+0.0001}$ & 13.08$_{-2.420}^{+1.687}$ & 87.207$_{-2.420}^{+1.687}$ & 0.009$_{-2.420}^{+1.687}$ & This work\\
\enddata
\end{deluxetable}

\begin{deluxetable}{l c c c c c c c}[!ht]
\tabletypesize{\footnotesize}
\tablecaption{BLS Properties of New Signal Candidates \label{tab:newplanets}}
\tablehead{
\colhead{TIC ID} & \colhead{RA} & \colhead{Dec} & \colhead{T$_0$ } & \colhead{Period} & \colhead{Depth} & \colhead{Disposition} &  \\
  &  \colhead{(J2000)} & \colhead{(J2000)}  &  \colhead{(BTJD-2457000)} & \colhead{(Days)} & & 
}
\startdata
234503282 & 00:46:22.92   &  -63:28:23.07   &  1336.1498 $\pm$ 0.0008 & 1.4376  $\pm$  0.0002  & 0.0171  $\pm$  0.0003 & Planet candidate\\  
234504626 & 00:47:45.69   &  -62:25:23.28   &  1335.5641 $\pm$ 0.0013 & 0.5341  $\pm$  0.0043  & 0.0035  $\pm$  0.0002 & Planet candidate\\
299780329 & 02:30:07.20   &  -79:45:23.23   &  1335.5707 $\pm$ 0.0022 & 1.6022  $\pm$  0.0074  & 0.0008  $\pm$  0.0001 & Planet candidate\\
394340319 & 02:37:27.78   &  -79:49:22.90   &  1336.3655 $\pm$ 0.0024 & 3.0371  $\pm$  0.0009  & 0.0068  $\pm$  0.0005 & Planet candidate\\  
350844139 & 06:00:07.22   &  -57:19:24.02   &  1335.5270 $\pm$ 0.0025 & 0.5709  $\pm$  0.0001  & 0.0038  $\pm$  0.0002 & Planet candidate\\
350930938 & 06:02:40.74   &  -54:50:10.85   &  1335.5272 $\pm$ 0.0026 & 2.6489  $\pm$  0.0013  & 0.0257  $\pm$  0.0009 & Planet candidate\\  
349155660 & 07:13:24.42   &  -63:59:18.69   &  1335.1362 $\pm$ 0.0014 & 0.69792 $\pm$  0.00002 & 0.0137  $\pm$  0.0006 & Planet candidate\\
349832804 & 07:34:34.61   &  -64:55:30.45   &  1335.2914 $\pm$ 0.0013 & 1.04140 $\pm$  0.00002 & 0.0102  $\pm$  0.0005 & Planet candidate\\
300810086 & 07:47:15.36   &  -69:01:50.52   &  1335.4133 $\pm$ 0.0016 & 0.66797 $\pm$  0.00007 & 0.0009  $\pm$  0.0004 & Planet candidate\\
159835004 & 21:21:00.62   &  -40:42:46.90   &  1335.4052 $\pm$ 0.0011 & 0.51606 $\pm$  0.00003 & 0.0058  $\pm$  0.0002 & Planet candidate\\
139771134 & 21:36:30.74   &  -52:30:46.85   &  1335.2118 $\pm$ 0.0011 & 0.91331 $\pm$  0.00002 & 0.0055  $\pm$  0.0003 & Planet candidate\\  
\hline
38813184  &  04:30:37.51  &  -62:16:01.48   &  1340.4600 $\pm$ 0.0022 & 5.3516  $\pm$ 0.0020  & 0.0123 $\pm$  0.0013 & Eclipsing binary\\  
231090180 &  04:35:54.59  &  -66:08:01.20   &  1335.5791 $\pm$ 0.0011 & 1.2545  $\pm$ 0.0002  & 0.0129 $\pm$  0.0007 & Eclipsing binary\\  
260003467 &  06:04:19.90  &  -57:18:09.50   &  1337.3377 $\pm$ 0.0016 & 2.6046  $\pm$ 0.0004  & 0.0043 $\pm$  0.0004 & Eclipsing binary\\  
349480507 &  07:23:44.80  &  -65:00:39.38   &  1335.2111 $\pm$ 0.0028 & 1.5628  $\pm$ 0.0004  & 0.1741 $\pm$  0.0075 & Eclipsing binary\\
349575582 &  07:27:56.92  &  -64:23:25.67   &  1335.8312 $\pm$ 0.0031 & 2.1383  $\pm$ 0.0008  & 0.0936 $\pm$  0.0050 & Eclipsing binary\\
350091587 &  07:40:30.45  &  -61:20:51.62   &  1335.9237 $\pm$ 0.0063 & 0.98006 $\pm$ 0.00006 & 0.0027 $\pm$  0.0003 & Eclipsing binary\\  
159834934 &  21:20:47.97  &  -40:54:39.78   &  1335.8908 $\pm$ 0.0059 & 2.7720  $\pm$ 0.0029  & 0.0182 $\pm$  0.0012 & Eclipsing binary\\
53896097  &  21:52:59.01  &  -24:47:50.20   &  1335.2516 $\pm$ 0.0022 & 0.74080 $\pm$ 0.00004 & 0.0111 $\pm$  0.0005 & Eclipsing binary\\  
301941187 &  21:57:14.13  &  -23:58:04.85   &  1337.0845 $\pm$ 0.0063 & 7.6676  $\pm$ 0.0014  & 0.0111 $\pm$  0.0017 & Eclipsing binary\\  
121490917 &  22:54:24.27  &  -46:38:39.17   &  1336.3423 $\pm$ 0.0020 & 2.4652  $\pm$ 0.0045  & 0.0478 $\pm$  0.0017 & Eclipsing binary\\
293525767 &  23:28:02.47  &  -73:38:44.76   &  1336.0060 $\pm$ 0.0034 & 6.8491  $\pm$ 0.0035  & 0.0290 $\pm$  0.0028 & Eclipsing binary\\  
\hline
234508527 & 00:50:16.11   &  -63:18:27.53   &  1335.18 & 1.231 & 0.173   & Contact binary\\
38907808  & 04:37:47.57   &  -65:16:33.97   &  1335.28 & 0.413 & 0.120   & Contact binary\\
349762067 & 07:30:33.93   &  -64:43:35.04   &  1335.31 & 0.858 & 0.029   & Contact Binary\\
29577927  & 20:58:45.47   &  -28:58:18.16   &  1335.15 & 0.904 & 0.117   & Contact binary\\
44625047  & 22:45:21.52   &  -46:49:41.88   &  1335.57 & 0.808 & 0.059   & Contact Binary\\  
261560715 & 23:16:07.06   &  -74:01:53.05   &  1335.13 & 0.692 & 0.049   & Contact binary\\
293507177 & 23:26:10.70   &  -73:23:49.92   &  1335.56 & 1.140 & 0.010   & Contact binary\\  
\hline
349647488 &  07:28:38.71  &  -64:20:56.37   &  1338.82 & 3.687 & 0.336   & Young stellar object\\
\hline
349647697 & 07:28:37.26   &  -64:38:11.58   &  --- & 1.104 & 0.011   & Stellar variability\\
29716907  & 21:02:13.27   &  -28:50:33.83   &  --- & 0.916 & 0.015   & Stellar variability\\
277874877 & 23:31:53.19   &  -73:01:33.11   &  --- & 1.521 & 0.010   & Stellar variability\\
\hline
234507163 &  00:50:00.62  &  -62:38:07.70   &  --- & 0.415 & 0.489   & RR Lyrae\\
29752683  &  21:03:16.64  &  -29:33:07.96   &  --- & 1.465 & 0.172   & RR Lyrae\\
293526531 &  23:29:26.71  &  -72:31:57.86   &  --- & 1.097 & 0.071   & RR Lyrae
\enddata
\tablecomments{The uncertainties from the depth are taken directly from the data; dilution from nearby sources would change the $R_p/R_\star$ of the presented signals.}
\end{deluxetable}

\end{document}